\documentclass[10pt,journal,compsoc]{IEEEtran}
\ifCLASSOPTIONcompsoc
  \usepackage[nocompress]{cite}
\else
  \usepackage{cite}
\fi
\ifCLASSINFOpdf
  \usepackage[pdftex]{graphicx}
\else
  \usepackage[dvips]{graphicx}
\fi
\usepackage{amsmath}
\usepackage{amsfonts,bm}
\usepackage{amssymb}
\usepackage{amsthm}
\usepackage{xcolor}

\usepackage{algorithmic}

\usepackage{array}

\ifCLASSOPTIONcompsoc
  \usepackage[caption=false,font=footnotesize,labelfont=sf,textfont=sf]{subfig}
\else
  \usepackage[caption=false,font=footnotesize]{subfig}
\fi
\usepackage{url}

\usepackage{algorithm}
\usepackage{multirow}
\usepackage{hyperref}
\usepackage{booktabs}

\begin{document}
%
\title{Spatial-Temporal Transformer Networks for Traffic Flow Forecasting}
%
%
%

\author{
 Mingxing~Xu, 
 Wenrui~Dai,~\IEEEmembership{Member,~IEEE,} 
 Chunmiao~Liu, 
 Xing~Gao, 
 Weiyao~Lin,~\IEEEmembership{Senior~Member,~IEEE,} 
 Guo-Jun~Qi,~\IEEEmembership{Senior~Member,~IEEE,} 
 and~Hongkai~Xiong,~\IEEEmembership{Senior~Member,~IEEE} 
}
\maketitle
\begin{abstract}
Traffic forecasting has emerged as a core component of intelligent transportation systems. However, it still remains an open challenge for timely accurate traffic forecasting, especially long-term forecasting, due to the highly nonlinear and dynamical spatial-temporal dependencies of traffic flows. In this paper, we propose a novel paradigm of Spatial-Temporal Transformer Networks (STTNs) that jointly leverage dynamical directed spatial dependencies and long-range temporal dependencies to improve the accuracy of long-term traffic flow forecasting. 
A new variant of graph neural networks, named spatial transformer, is presented to dynamically model directed spatial dependencies with self-attention mechanism to capture real-time conditions and directions of traffic flows. Various patterns of spatial dependencies are jointly modeled with multi-head attention mechanism to consider multiple factors, including similarity, connectivity and covariance. Furthermore, temporal transformer is developed to model long-range bidirectional temporal dependencies across multiple time steps. 
In comparison to existing works, STTNs enable efficient and scalable training for long-range spatial-temporal dependencies. Experimental results demonstrate that STTNs are competitive with the state-of-the-arts, especially for long-term traffic flow forecasting, on real-world PeMS-Bay and PeMSD7(M) datasets.
\end{abstract}


%

\section{Introduction}
\IEEEPARstart{W}{ith} the deployment of affordable traffic sensor technologies, the exploding traffic data are bringing us to the era of transportation big data. Intelligent Transportation System (ITS)~\cite{mori2015review} is thus developed to leverage transportation big data for efficient urban traffic controlling and planning. As a core component of ITS, accurate traffic forecasting in a timely fashion has attracted increasing attentions.

In traffic forecasting, the future traffic conditions (e.g. speeds, volumes and density) of a node are predicted from the historical traffic data of itself and its neighboring nodes. It is important for a forecasting model to effectively and efficiently capture the spatial and temporal dependencies within traffic flows. Traffic forecasting is commonly classified into two scales, i.e., short-term ($\le$30 minutes) and long-term ($\ge$30 minutes). Existing approaches like time-series models~\cite{liu2011discovering} and Kalman filtering ~\cite{lippi2013short} perform well on short-term forecasting. However, the stationary assumption for these models is not practical in long-term forecasting, as traffic flows are highly dynamical in nature. Furthermore, they fail to jointly exploit the spatial and temporal correlations in the traffic flows to make long-term forecasting. 

Traffic networks can be represented as graphs in which the nodes represent traffic sensors and the edges together with their weights are determined by the connectivity as well as Euclidean distances between sensors. Consequently, traffic flows can be viewed as graph signals evolving with time. Recently, Graph Neural Networks (GNNs) \cite{atwood2016diffusion,defferrard2016convolutional,kipf2017semi} have emerged as a powerful tool for processing graph data. Sequential models are improved by incorporating GNNs to jointly capture spatio-temporal correlations for both short-term and long-term forecasting. GNN-based traffic forecasting models are first developed in \cite{yu2018spatio} and \cite{li2018dcrnn_traffic} to improve the prediction performance by introducing the inherent topology of traffic networks into sequential models. Spatial \cite{atwood2016diffusion} or spectral \cite{defferrard2016convolutional} Graph Convolutional Networks (GCNs) are integrated with convolution-based sequence learning models \cite{gehring2017convolutional} or recurrent neural networks (RNNs) to jointly capture the spatial and temporal dependencies.
However, these models are still restricted for traffic forecasting, especially long-term prediction, in the following two aspects:

\begin{figure}[!t]
\renewcommand{\baselinestretch}{1.0}
\centering
\subfloat[]{\includegraphics[width=0.48\textwidth]{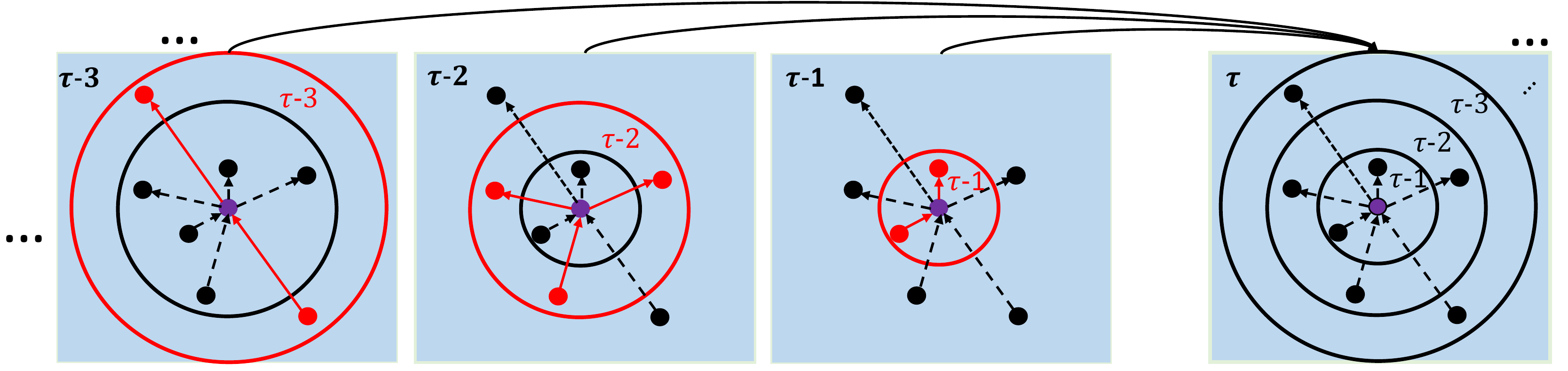}}\\
\subfloat[]{\includegraphics[width=0.48\textwidth]{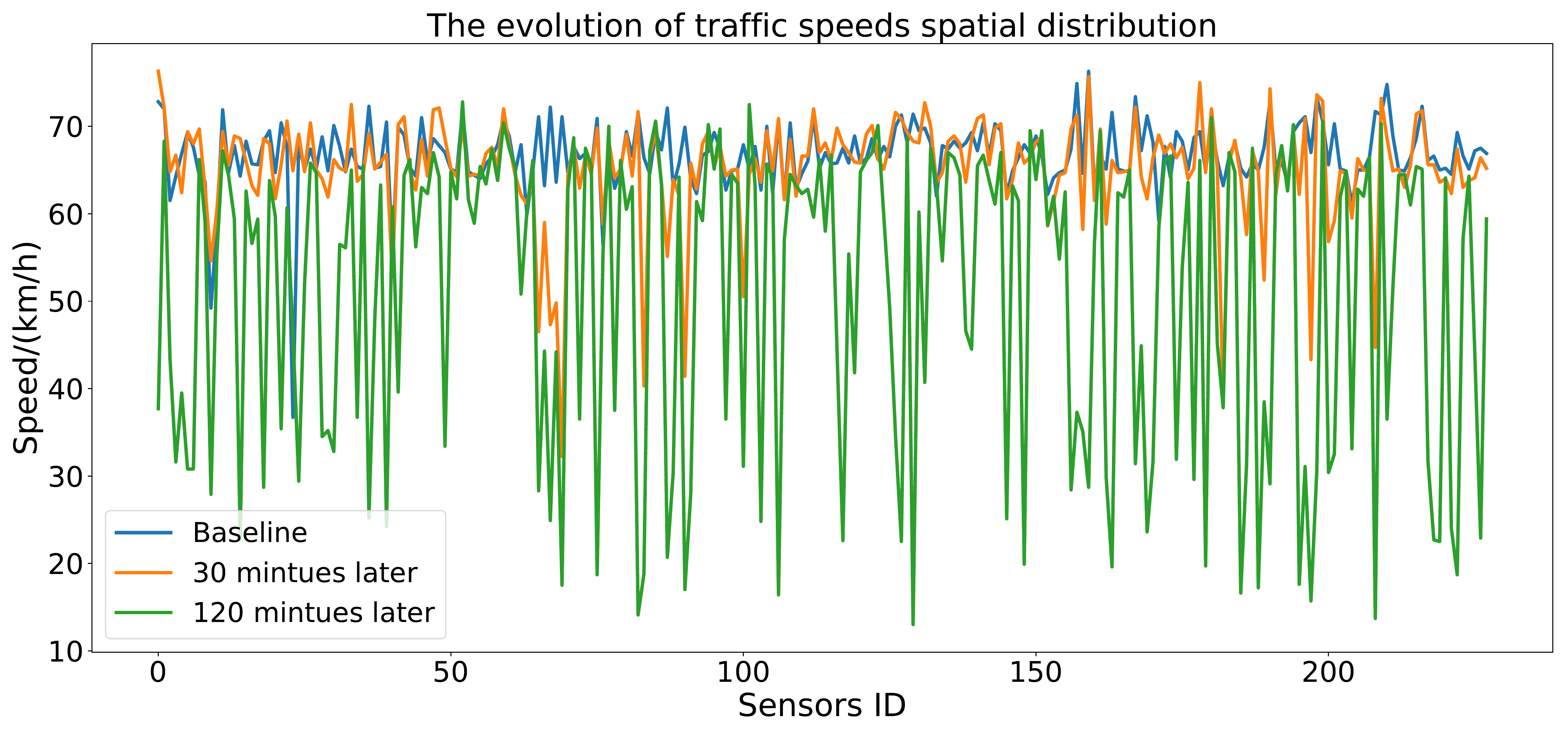}}\\
\caption{(a) Traffic forecasting models with joint spatial temporal dependencies where spatial dependencies are evolving with time. (b) Evolution of the spatial distribution of real-time traffic speeds.}\label{fig0}
\end{figure}

\begin{figure*}[!t]
 \renewcommand{\baselinestretch}{1.0}
 \centering
 \subfloat[Autoregressive short-term traffic prediction]{\includegraphics[width=0.45\textwidth]{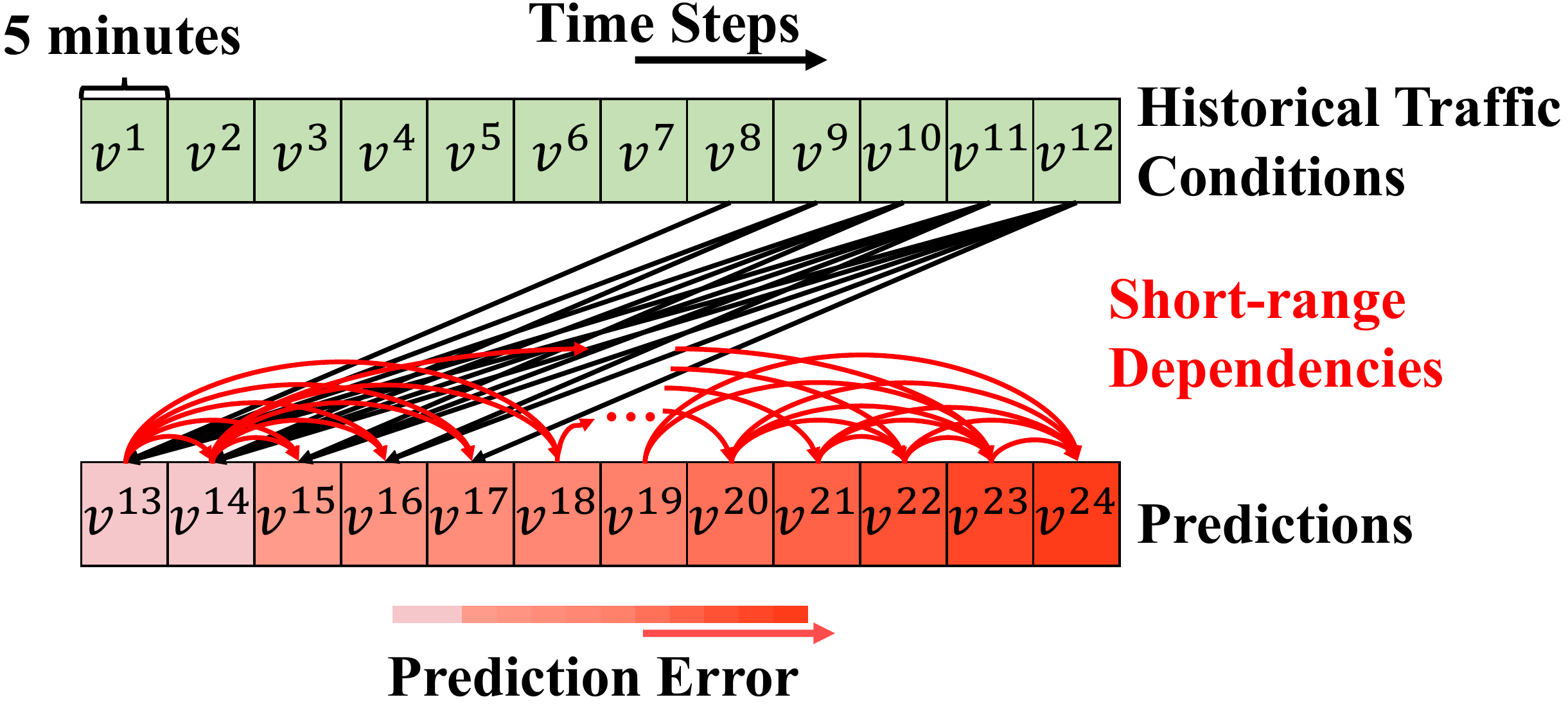}}
 \subfloat[Multi-step long-term traffic prediction]{\includegraphics[width=0.45\textwidth]{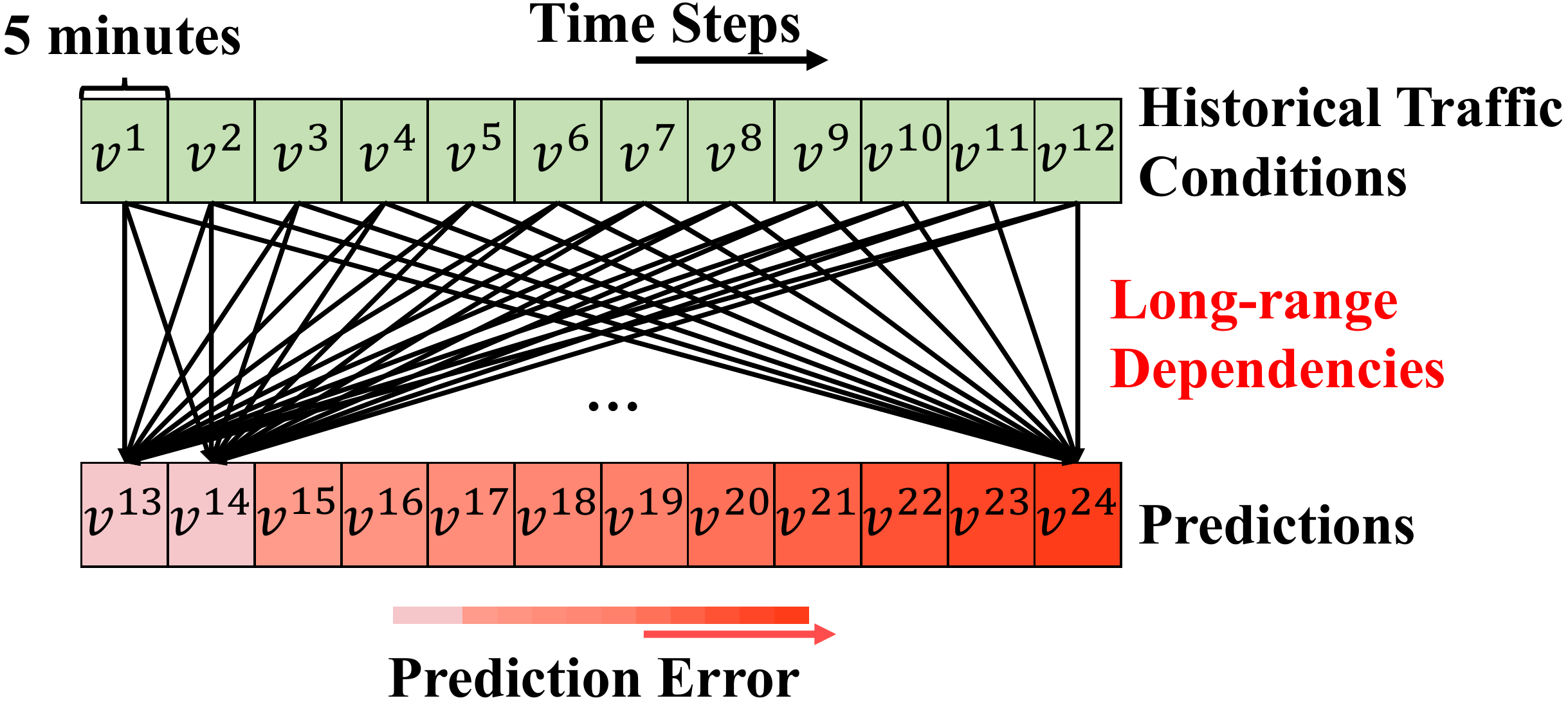}}\\
 \caption{Autoregressive short-term prediction with short-range temporal dependencies and multi-step long-term prediction with long-range temporal dependencies for traffic flow forecasting. Traffic data are aggregated every five minutes. Green rectangles are error-free historical observations, while red ones are predictions. For predictions (red), darker color stands for larger prediction error.
 }
\label{fig1}
\end{figure*}

{\noindent \bf Fixed Spatial Dependencies:} 
In traffic forecasting task, spatial dependencies are highly dynamical due to 
road topology, varying traffic speeds and multiple factors (e.g., weather conditions, rush hours and traffic accidents). For each sensor, its correlated sensors vary with the time steps.
Fig.~\ref{fig0}(a) provides a simple example for traffic forecasting, where traffic speeds excluding target nodes (purple) do not change across different time steps. For the target node (purple), different correlated nodes (red) are considered to formulate the spatial dependencies for different time steps (e.g., $\tau-1$, $\tau-2$, $\tau-3$), according to the range (red circle) determined by the traffic speeds and distances. Given the connectivity and distance between arbitrary two sensors, their spatial dependencies are complicated, due to the time-varying traffic speeds. Spatial dependencies would also vary for traffic flows with different directions, i.e., upstream and downstream. Furthermore, spatial dependencies irregularly oscillate with time steps, due to the periodical effect of rush hours, varying weather conditions and unexpected occurrence of traffic accidents, as shown in Fig.~\ref{fig0}(b). Consequently, it is necessary to effectively capture these dynamical spatial dependencies to improve traffic forecasting.

{\noindent \bf Limited-range Temporal Dependencies:} Long-range temporal dependencies are usually ignored in existing methods. However, long-term traffic flow forecasting can be facilitated by considering dependencies with varying scales for different time steps. 

Fig.~\ref{fig0}(a) illustrates spatial dependencies with various scales for different time steps, which implies that  prediction performance would be degraded by limiting the range of temporal dependencies. Furthermore, prediction errors would be propagated and accumulated for long-term traffic forecasting with existing auto-regressive methods trained with either individual loss for each time step~\cite{yu2018spatio} or joint loss for multiple time steps~\cite{li2018dcrnn_traffic}, as depicted in Fig.~\ref{fig1}(a). 

It is desirable to achieve accurate long-term prediction based on long-range temporal dependencies extracted from error-free temporal contexts, as shown in Fig.~\ref{fig1}(b). 

In this paper, we propose a novel paradigm of Spatial-Temporal Transformer Networks (STTNs) to address aforementioned challenges in traffic flow forecasting. The contributions of this paper are summarized as below.
\begin{itemize}
\item We develop a spatial-temporal block to dynamically model long-range spatial-temporal dependencies.
\item We present a new variant of GNNs, named {\em spatial transformer}, to model the time-varying directed spatial dependencies and dynamically capture the hidden spatial patterns of traffic flows.
\item We design a {\em temporal transformer} to achieve multi-step prediction using long-range temporal dependencies.
\end{itemize}

To be concrete, the spatial transformer dynamically models directed spatial dependencies based on the real-time traffic speeds, connectivity and distance between sensors and directions of traffic flows. High-dimensional latent subspaces are learned from the input spatial-temporal features together with positional embeddings of road topology and temporal information to infer time-varying spatial dependencies. 
To represent abrupt changes of traffic flows, long-range time-varying hidden patterns are captured from local and global dependencies using the self-attention mechanism. Furthermore, the temporal transformer simultaneously achieves multi-step predictions for future traffic conditions based on the long-range temporal dependencies. It suppresses the propagation of prediction error and allows parallel training and prediction to improve efficiency, as illustrated in Fig.~\ref{fig1}.  


Different from~\cite{wu2019graph}, STTN facilitates traffic flow forecasting by dynamically modeling sptail-temporal dependencies varying with road topology and time steps, rather than fixed spatial dependencies. To validate the efficacy of STTN, we evaluate it on two real-world traffic datasets, i.e., PeMSD7(M) and PEMS-BAY. Extensive experiments demonstrate that STTNs can achieve  the state-of-the-arts performance in traffic flow forecasting, especially for long-term predictions.

The rest of this paper is organized as follows. In Section~\ref{sec:related}, we briefly review existing approaches for modeling spatial and temporal dependencies. Section~\ref{sec:model} formulates the spatial-temporal graph prediction problem for traffic flow forecasting and elaborates the proposed STTN for solution. Extensive experiments on real-world traffic datasets are performed in Section~\ref{sec:exp} to evaluate STTN with the state-of-the-art methods. Finally, we conclude this paper and discuss the further work in Section~\ref{sec:con}.

\section{Related Work}\label{sec:related}
We begin with a brief overview of the existing approaches for modeling spatial and temporal dependencies in traffic flow forecasting.

\subsection{Spatial Dependencies}
Statistical and neural network based models are first developed for traffic flow forecasting. 
Statistical models like autoregressive integrated moving average (ARIMA)~\cite{min2011real} and Bayesian networks~\cite{wang2014new} model spatial dependencies from a probabilistic view. Although they help to analyze the uncertainty within traffic flows, their linear natures impedes them to effectively model the highly-nonlinearity within traffic flows. 
Neural networks are introduced to capture the nonlinearity of traffic flows, but their fully-connected structures are computation intensive and memory consuming. 
Furthermore, the lack of assumptions make it impossible to capture the complicated spatial patterns in traffic flows.

With the development of convolutional neural networks (CNNs), they have been employed for traffic forecasting considering their powerful feature extraction abilities in many applications~\cite{gehring2017convolutional,Krizhevsky:2012:ICD:2999134.2999257,long2015fully}. CNNs are adopted in~\cite{ma2017learning,zhang2017deep,zhang2019flow} to extract the spatial features in which traffic networks are converted to regular grids. However, this grid conversion leads to the loss of inherent topology information characterizing irregular traffic networks. Graph neural networks (GNNs) \cite{scarselli2008graph,gilmer2017neural} are developed to generalize the deep learning to non-Euclidean domains. As a variant of GNNs, graph convolution networks (GCNs)~\cite{atwood2016diffusion,defferrard2016convolutional,kipf2017semi} generalize classical convolutions to the graph domain. Recently, GCNs are widely considered to model the spatial dependencies of traffic flows to explore the inherent traffic topology. STGCN \cite{yu2018spatio} models spatial dependencies with spectral graph convolutions defined on an undirected graph, while DCRNN \cite{li2018dcrnn_traffic} employs diffusion graph convolutions on a directed graph to accommodate the directions of traffic flows. However, they ignore the dynamic changes of traffic conditions (e.g. rush hours and traffic accidents), as spatial dependencies are fixed once trained. In \cite{guo2019attention}, spatial dependencies are dynamically generated with the depth of spatial and temporal blocks, rather than the actual time steps. Dynamical spatial dependencies are modeled by incorporating the graph attention networks (GATs) \cite{velivckovic2017graph} and the embedded geo-graph features summarized by extra meta-learner \cite{pan2019urban}. However, the predefined graph topology using $k$ nearest neighbors is limited to discover hidden patterns of spatial dependencies at various scales beyond local nodes. Graph WaveNet \cite{wu2019graph} improves the accuracy of traffic forecasting with hidden spatial patterns through a learnable embedding for each node in the graph, but their spatial dependencies are still fixed once trained. In this paper, STTNs efficiently model dynamical directed spatial dependencies in high-dimensional latent subspaces, rather than adopt predefined graph structures and local nodes.

\begin{figure*}[!t]
\renewcommand{\baselinestretch}{1.0}
\centering
\includegraphics[width=0.9\textwidth]{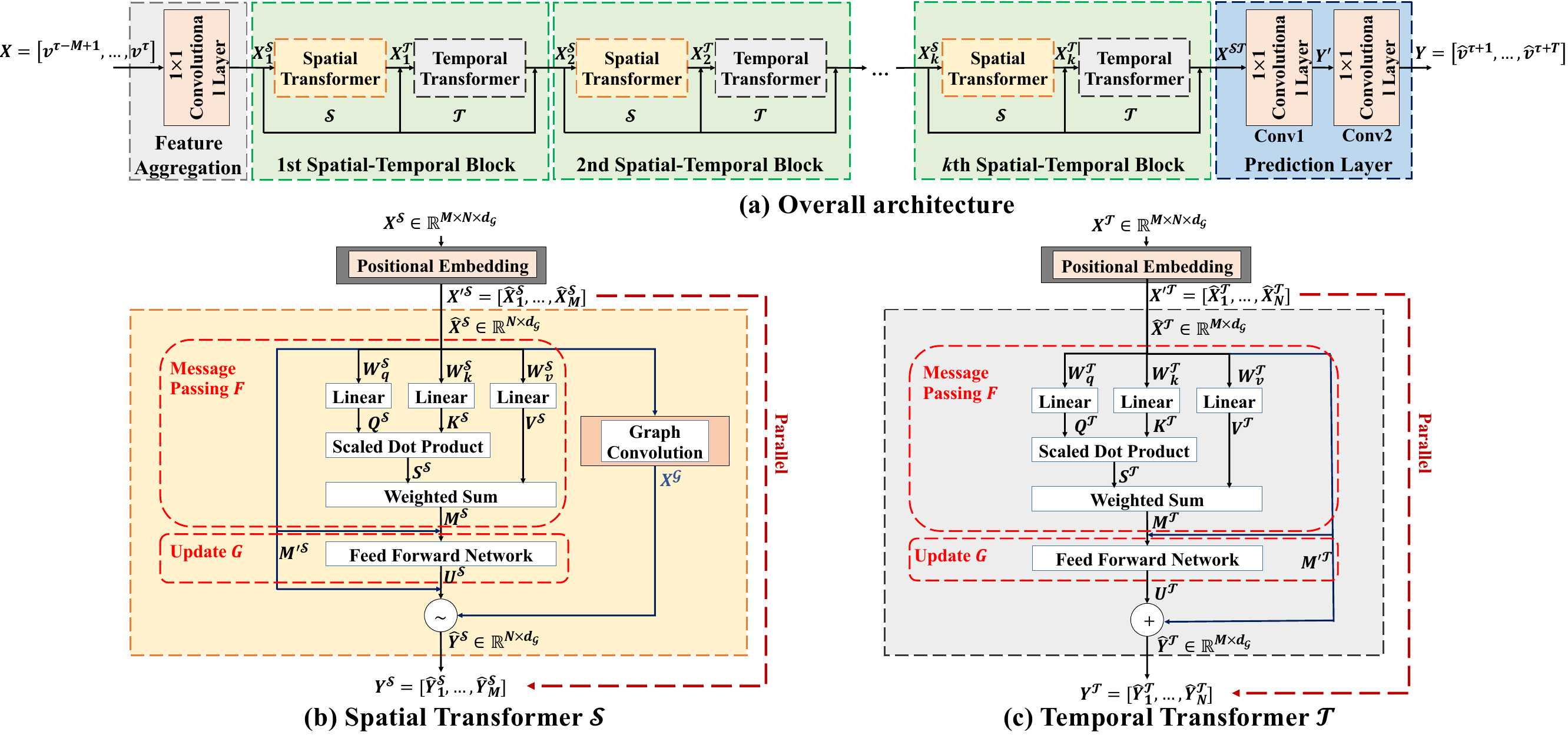}
\caption{Illustrative architecture of the proposed spatial-temporal transformer network (STTN). STTN consists of stacked spatial-temporal (ST) blocks and one prediction layer. Each ST block leverages one spatial transformer and one temporal transformer to jointly model the spatial-temporal dependencies. Skip connections are adopted to combine all levels of spatial-temporal features by the ST blocks.
}\label{fig2}
\end{figure*}

\subsection{Temporal Dependencies}
As stated in \cite{ma2015long} and \cite{wu2016short}, RNNs are limited for modeling temporal dependencies, due to exploding or vanishing gradients in training and inaccurate determination of sequence lengths.  To alleviate these drawbacks, Gated Recurrent Units (GRUs) \cite{chung2014empirical} and Long-Short Term Memory (LSTM) \cite{Hochreiter:1997:LSM:1246443.1246450} are developed to model long-range dependencies for traffic forecasting \cite{pan2019urban,guo2019attention,ma2015long,wu2016short} However, these sequential models still suffer from time-consuming training process and limited scalability for modeling long sequences. Convolution-based sequence learning models \cite{gehring2017convolutional} are adopted as an alternative \cite{yu2018spatio,guo2019attention}, but require multiple hidden layers to cover large contexts under limited size of receptive fields. WaveNet with dilation convolution is adopted in \cite{wu2019graph} to enlarge the receptive fields, and consequently, reduce the number of hidden layers. However, its model scalability is restricted for long input sequences, as the number of hidden layers increases linearly with the lengths of input sequences. Furthermore, the efficiency for capturing long-range dependencies would be affected by deeper layers, due to the increasing lengths of paths between components in the sequence \cite{NIPS2017_7181,hochreiter2001gradient}. These facts imply that it would be prohibitive to find the optimal lengths of input sequences, as the model needs to be redesigned for input sequences with different lengths. Transformers \cite{NIPS2017_7181} achieve efficient sequence learning with the highly parallelizable self-attention mechanism. Long-range time-varying dependencies can be adaptively captured from input sequences with various lengths with one single layer.

\section{Proposed Model}\label{sec:model}
In this section, we introduce the proposed spatial-temporal transformer network. We first formulate the traffic forecasting task as a spatial-temporal prediction problem. To address this problem, we describe the overall architecture of the proposed model that consists of two main components: spatial-temporal (ST) block and prediction layer. Subsequently, we elaborate the proposed {\em spatial transformer} and {\em temporal transformer} in ST block, respectively.

\subsection{Problem Formulation}\label{sub:formulation}
A traffic network can be naturally represented as a graph ${\mathcal G}=({\mathcal V, \mathcal E }, A)$, where ${\mathcal V}$ is the set of $N$ nodes representing the sensors, ${\mathcal E}$ is the set of edges reflecting physical connectivity between sensors and $A\in {\mathbb R^{N\times N}}$ is the adjacency matrix constructed with the Euclidean distances between sensors via Gaussian kernel. Traffic forecasting is a classic spatial-temporal prediction problem. In this paper, we focus on forecasting traffic speeds $v^\tau\in {\mathbb R^N}$ at time step $\tau$ for the $N$ sensors, and volume and density can be similarly calculated as traffic speeds. Given $M$ historical traffic conditions $[v^{\tau-M+1},\cdots, v^\tau]$ observed by the $N$ sensors and a traffic network $\mathcal G$, a traffic forecasting model ${\mathcal F}$ is learned to predict $T$ future traffic conditions $[{\hat v}^{\tau+1},\cdots,{\hat v}^{\tau+T}]$. 
\begin{equation}\label{eq1}
{\hat v}^{\tau+1},\cdots,{\hat v}^{\tau+T}={\mathcal F}(v^{\tau-M+1},\cdots, v^\tau;{\mathcal G})
\end{equation}
To achieve accurate prediction, $\mathcal F$ captures dynamical spatial dependencies $S_\tau^\mathcal{S}\in {\mathbb R^{N\times N}}$ and long-range temporal dependencies $S_\tau^\mathcal{T}\in {\mathbb R^{M\times M}}$ from $v^{\tau-M+1},\cdots,v^\tau$ and $\mathcal G$. However, existing methods are limited in long-term prediction, as they only consider fixed spatial dependencies and short-range temporal dependencies. In this paper, spatial transformer is developed to dynamically train $\mathcal F$ with time-varying spatial dependencies $S^{\mathcal S}_\tau\in {\mathbb R^{N\times N}}$ for each time step $\tau$, as defined in Section~\ref{sub:spatial}. Furthermore, long-term temporal dependencies $S_\tau^\mathcal{T}\in {\mathbb R^{M\times M}}$ are efficiently learned with the temporal transformer based on self-attention mechanism, as shown in Section~\ref{sub:temporal}.
Error propagation can be addressed by simultaneously making $T$ predictions ${\hat v}^{\tau+1},\cdots,{\hat v}^{\tau+T}$ from the error-free historical observations $v^{\tau-M+1},\cdots,v^\tau$ with the spatial-temporal features learned based on $S_\tau^\mathcal{S}$ and $S_\tau^\mathcal{T}$. 
For simplicity, we omit the subscript $\tau$ in $S_\tau^\mathcal{S}$ and $S_\tau^\mathcal{T}$ in the rest of this section.

\subsection{Overall Architecture}\label{sub:overall}
As depicted in Fig.~\ref{fig2}, the proposed spatial-temporal transformer network consists of stacked spatial-temporal blocks and a prediction layer. Here, each spatial-temporal block is composed of one spatial transformer and one temporal transformer to jointly extract spatial-temporal features in the context of dynamical dependencies. Spatial-temporal blocks can be further stacked to form deep models for deep spatial-temporal features. Subsequently, the prediction layer utilizes two $1\times 1$ convolutional layers to aggregate these spatial-temporal features for traffic forecasting. 

\subsubsection{Spatial-temporal Blocks}
The future traffic conditions of one node are determined by the traffic conditions of its neighboring nodes, the time steps of observations and abrupt changes like traffic accidents and weather conditions. In this section, we develop a spatial-temporal block to integrate spatial and temporal transformers to jointly model the spatial and temporal dependencies within traffic networks for accurate prediction, as illustrated in Fig.~\ref{fig2}.
The input to the $l$-th spatial-temporal block is a 3-D tensor $X_l^{\mathcal S}\in {\mathbb R}^{M\times N\times d_{\mathcal G}}$ of $d_{\mathcal G}$-dimensional features for the $N$ nodes at time steps $\tau-M+1,\cdots,\tau$ extracted by the $l-1$-th spatial-temporal block. The spatial transformer $\mathcal S$ and temporal transformer ${\mathcal T}$ are stacked to generate the 3-D output tensor. Residual connections are adopted to for stable training. In the $l$-th spatial-temporal block, the spatial transformer $\mathcal{S}$ extracts spatial features $Y_l^{\mathcal S}$ from the input node feature $X_l^{\mathcal S}$ as well as graph adjacency matrix $A$. 
\begin{equation}\label{eq:spatial}
 Y_l^{\mathcal S}={\mathcal S}(X_l^{\mathcal S},A)
\end{equation}
$Y_l^{\mathcal S}$ is combined with $X_l^{\mathcal S}$ to generate the input $X_l^{\mathcal T}$ to the subsequent temporal transformer. 
\begin{equation}\label{eq:temporal}
 Y_l^{\mathcal T}={\mathcal T}(X_l^{\mathcal T})    
\end{equation}
Consequently, we obtain the output tensor $X_{l+1}^{\mathcal S}=Y_l^{\mathcal T}+X_l^{\mathcal T}$ and feed $X_{l+1}^{\mathcal S}$ into the $l+1$-th spatial-temporal block. 
Multiple spatial-temporal blocks can be stacked to improve the model capacity according to the tasks at hand. In Section~\ref{sub:spatial} and \ref{sub:temporal}, we elaborate the spatial and temporal transformers. Without loss of generality, we omit the subscript $l$ of $X_l^{\mathcal S}$, $X_l^{\mathcal T}$, $Y_l^{\mathcal S}$ and $Y_l^{\mathcal T}$ for the $l$-th spatial-temporal block.

\subsubsection{Prediction Layer}
The prediction layer leverages two classical convolutional layers to make multi-step prediction based on the spatial-temporal features from the last spatial-temporal block. Its input is a 2-D tensor $X^{\mathcal {ST}}\in {\mathbb R}^{N\times d^{{\mathcal {ST}}}}$ that consists of the $d^{{\mathcal {ST}}}$-dimensional spatial-temporal features of the $N$ nodes for last time step $\tau$.
The multi-step prediction $Y\in {\mathbb R}^{N\times T}$ for $T$ future traffic conditions of the $N$ nodes is
\begin{equation}
Y=\text{Conv}(\text{Conv}(X^{\mathcal {ST}}))
\end{equation}
Mean absolute loss are adopted to train the model.
\begin{equation}\label{eqloss}
L=\|Y-Y^{gt}\|_1
\end{equation}
where $Y^{gt}\in {\mathbb R}^{N\times T}$ is the groundtruth traffic speeds.

\subsection{Spatial Transformer}\label{sub:spatial}
As shown in Fig.~\ref{fig2}(b), the spatial transformer consists of spatial-temporal positional embedding layer, fixed graph convolution layer, dynamical graph convolution layer and gate mechanism for information fusion. The spatial-temporal positional embedding layer incorporates spatial-temporal position information (e.g., topology, connectivity, time steps) into each node. According to~\cite{diao2019dynamic}, the traffic signal over a period of time can be decomposed into a stationary component determined by the road topology (e.g., connectivity and distance between sensors) and a dynamical component determined by real-time traffic conditions and sudden changes (e.g., accidents and weather changes). Consequently, we develop a fixed graph convolutional layer and a dynamical graph convolutional layer to explore the stationary and directed dynamical components of spatial dependencies, respectively. The learned stationary and dynamical spatial features are fused with gate mechanism. We further show that the proposed spatial transformer can be viewed as a general message passing GNN for dynamical graph construction and feature learning. 

\subsubsection{Spatial-Temporal Positional Embedding}
Fig.~\ref{fig0}(a) shows that the spatial dependencies of two nodes in the graph would be determined by their distances and observed time steps. Transformer~\cite{NIPS2017_7181} cannot capture the spatial (position) and temporal information of observations with the fully connected feed-forward structures. 
Thus, the prior positional embedding is required to inject the 'position' information into the input sequences. 
In the proposed spatial transformer, we adopt learnable spatial and temporal positional embedding layer to learn the spatial-temporal embedding into each node feature. The dictionaries ${\mathcal {\hat D}}^{\mathcal S}\in\mathbb{R}^{N\times N}$ and ${\mathcal {\hat D}}^{\mathcal T}\in \mathbb{R}^{M\times M}$ are learned as spatial and temporal positional embedding matrices, respectively. ${\mathcal {\hat D}}^{\mathcal S}$ is initialized with the graph adjacency matrix to consider the connectivity and distance between nodes for modeling spatial dependencies, while ${\mathcal {\hat D}}^{\mathcal T}$ is initialized with one-hot time encoding to inject the time step into each node. ${\mathcal {\hat D}}^{\mathcal S}$ and ${\mathcal {\hat D}}^{\mathcal T}$ are tiled along the spatial and temporal axes to generate ${\mathcal D}^{\mathcal S}\in {\mathbb R}^{M\times N \times N}$ and ${\mathcal D}^{\mathcal T}\in {\mathbb R}^{M\times N \times M}$, respectively. the embedded features $X'^{\mathcal S}=F_t([X^{\mathcal S}, {\mathcal D}^{\mathcal S}, {\mathcal D}^{\mathcal T}])\in\mathbb{R}^{M\times N\times d_{\mathcal G}}$ with a fixed dimension of $d_{\mathcal G}$ are obtained from $X^{\mathcal S}\in {\mathbb R}^{M\times N \times d_{\mathcal G}}$ that concatenates ${\mathcal D}^{\mathcal S}\in {\mathbb R}^{M\times N \times N}$ and ${\mathcal D}^{\mathcal T}\in {\mathbb R}^{M\times N \times M}$. Here, $F_t$ is a $1\times 1$ convolutional layer to transform the concatenated features into a $d_{\mathcal G}$-dimensional vector for each node at each time step. $X'^{\mathcal S}$ is fed into the fixed and dynamical graph convolutional layers for spatial feature learning. Since graph convolution operations can be realized in parallel for the $M$ time steps via tensor operations, we consider the 2-D tensor $\hat X^{\mathcal S}\in {\mathbb R}^{N\times d_{\mathcal G}}$ of $X'^{\mathcal S}$ for arbitrary one time step for brevity.

\subsubsection{Fixed Graph Convolutional Layer}\label{sub:fconv}
Graph convolution is generalizes classical grid-based convolution to the graph domain. Node features are derived by aggregating the information from its neighboring nodes based on the learned weights and predefined graph. 
In this subsection, graph convolution based on Chebyshev polynomial approximation is employed to learn the structure-aware node features, and consequently, capture the stationary spatial dependencies from the road topology. 
Let us denote $D$ the degree matrix of $\mathcal{G}$ with its diagonal elements $D_{ii}=\sum_i{A_{ij}}$ for $i=1,\cdots,N$. The normalized Laplacian matrix $L$ is defined by $L=I_n-D^{-1/2}AD^{-1/2}$ and the scaled Laplacian matrix ${\widetilde L}=2L/\lambda_{max} -I_n$ for Chebyshev polynomials, where $\lambda_{max}$ is the largest eigenvalues of $L$. Given the embedded features ${\hat X}^{\mathcal S}$, the structure-aware node features ${\hat X}^{\mathcal G}\in {\mathbb R}^{N\times d_{\mathcal G}}$ are obtained with the graph convolution approximated by the Chebyshev polynomials $T_k$ with the orders $k=1,\cdots,K$ for each time step.
\begin{equation}
\hat X^{{\mathcal G}}_{:,j}=\sum\limits_{i=1}^{d_{\mathcal G}}\sum\limits_{k=0}^K \theta_{ij,k}T_k({\widetilde L})\hat X^{\mathcal S}_{:,i} \quad \forall j=0,\cdots,d_{\mathcal G},
\end{equation}
where $\hat X^{\mathcal G}_{:,j}$ is the $j$-th channel (column) of $\hat X^{{\mathcal G}}$ and $\theta_{ij,k}$ is the learned weights. 
Since $\mathcal{G}$ is constructed based on the physical connectivity and distance between sensors, the stationary spatial dependencies determined by the road topology can be explicitly explored through the fixed graph convolutional layer.

\subsubsection{Dynamical Graph Convolutional Layer}
GCN-based models like \cite{yu2018spatio} and \cite{li2018dcrnn_traffic} only model stationary spatial dependencies. To capture the time-evolving hidden spatial dependencies, we propose a novel dynamical graph convolutional layer to achieve training and modeling in the high-dimensional latent subspaces. Specifically, we learn the linear mappings that project the input features of each node to the high-dimensional latent subspaces. As shonw in Fig.~\ref{fig3}, the self-attention mechanism is adopted for the projected features to efficiently model the dynamical spatial dependencies between nodes according to the varying graph signals. 
Although it is also adopted in \cite{guo2019attention}, the weights of edges are calculated based on the predefined road topology. Predefined road topology cannot sufficiently represent the dynamical spatial dependencies within traffic networks. Consequently, we learn multiple linear mappings to model dynamical directed spatial dependencies affected by various factors in various latent subspaces.


The embedded feature $\hat X^{\mathcal S}$ of each time step is first projected into the high-dimensional latent subspaces. The mappings are realized with the feed-forward neural networks. When the single-head attention model is considered for one pattern of spatial dependencies, three latent subspaces are trained for each node based on $\hat X^{\mathcal S}$, including the query subspace spanned by $Q^{\mathcal S}\in {\mathbb R}^{N\times d^{\mathcal S}_{\mathcal A}}$, the key subspace by $K^{\mathcal S}\in {\mathbb R}^{N\times d^{\mathcal S}_{\mathcal A}}$ and the value subspace by $V^{\mathcal S}\in {\mathbb R}^{N\times d_{\mathcal G}}$. 
\begin{align}\label{eqa}
 Q^{\mathcal S}&=\hat X^{\mathcal S}W^{\mathcal S}_q\nonumber\\
 K^{\mathcal S}&=\hat X^{\mathcal S}W^{\mathcal S}_k\nonumber\\
 V^{\mathcal S}&=\hat X^{\mathcal S}W^{\mathcal S}_v
\end{align}
Here, $W^{\mathcal S}_q\in\mathbb{R}^{d_{\mathcal G}\times d_{\mathcal A}^\mathcal{S}}$, $W^{\mathcal S}_k\in\mathbb{R}^{d_{\mathcal G}\times d_{\mathcal A}^\mathcal{S}}$ and $W^{\mathcal S}_v\in\mathbb{R}^{d_{\mathcal G}\times d_{\mathcal G}}$ are the weight matrices for $Q^{\mathcal S}$, $K^{\mathcal S}$ and $V^{\mathcal S}$, respectively.

Dynamical spatial dependencies $S^{\mathcal S}\in {\mathbb R^{N\times N}}$ between nodes are calculated with the dot product of $Q^{\mathcal S}$ and $K^{\mathcal S}$.
\begin{equation}\label{eq4}
S^{\mathcal S}=\text{softmax}(Q^{\mathcal S}(K^{\mathcal S})^T/\sqrt{d^{\mathcal S}_{\mathcal A}})
\end{equation}
In Eq.~\eqref{eq4}, dot product is adopted reduce the computational and storage costs in calculation. Softmax is used to normalize the spatial dependencies and the scale $\sqrt{d^{\mathcal S}_{\mathcal A}}$ prevents the saturation led by Softmax function.  
Thus,the node features $ M^\mathcal{S}\in{\mathbb R^{N\times d_{\mathcal G}}}$ are updated with $S^{\mathcal S}$.
\begin{equation}\label{eq5}
M^\mathcal{S}=S^{\mathcal S}V_{\mathcal S}.
\end{equation}
It is worth mentioning that multiple patterns of spatial dependencies can be learned with multi-head attention mechanism by introducing multiple pairs of subspaces, which is able to model different hidden spatial dependencies from various latent subspaces.

Furthermore, a shared three-layer feed-forward neural network with nonlinear activation is applied on each node to further improve the prediction conditioned on the learned node features $M^\mathcal{S}$. The interactions among feature channels are explored to update $M^\mathcal{S}$ with $U^{\mathcal S}\in {\mathbb R}^{N \times d_{\mathcal G}}$.
\begin{equation}\label{eq6}
U^{\mathcal S}=\text{ReLu}(\text{ReLu}(\hat M'^{\mathcal S}W^{\mathcal S}_0)W^{\mathcal S}_1)W^{\mathcal S}_2
\end{equation}
where $M'^{\mathcal S}=\hat X^{\mathcal S}+M^{\mathcal S}$ is the residual connection for stable training and $W^{\mathcal S}_0$, $W^{\mathcal S}_1$ and $W^{\mathcal S}_2$ are the weight matrices for the three layers. $U^{\mathcal S}$ and $M'^{\mathcal S}$ are combined by $\hat Y^{\mathcal S}=U^{\mathcal S}+M'^{\mathcal S}$ for feature fusion with the gate mechanism. It should be noted that we can stack multiple dynamical graph convolution layers for deep models to improve the model capacity for complicated spatial dependencies. 

\subsubsection{Gate Mechanism for Feature Fusion}
The gate mechanism is developed to fuse the spatial features learned from fixed and dynamical graph convolutional layers. The gate $g$ is derived from $Y'^{\mathcal S}$ and $X^{\mathcal S}$ of the fixed and dynamical graph convolutional layers.
\begin{equation}
g=\text{sigmoid}(f_{\mathcal S}(\hat Y^{\mathcal S})+f_{\mathcal G}(X^{\mathcal G})),
\end{equation}
where $f_{\mathcal S}$ and $f_{\mathcal G}$ are linear projections that transform $Y'^{\mathcal S}$ and $X^{\mathcal G}$ into 1-D vectors, respectively. As a result, the output $Y^{\mathcal S}$ is obtained by weighting $Y'^{\mathcal S}$ and $X^{\mathcal S}$ with the gate $g$.
\begin{equation}
Y'^{\mathcal S}=g\hat Y^{\mathcal S}+(1-g)X^{\mathcal G}
\end{equation}
The output $Y^{\mathcal S} \in {\mathbb R}^{M\times N\times d_{\mathcal G}} $ of the spatial transformer collects $Y'^{\mathcal S}$ obtained in parallel for the $M$ time steps and is fed into the subsequent temporal transformer with $X^{\mathcal T}=Y^{\mathcal S}$. 
\subsubsection{General Dynamical Graph Neural Networks}
Existing spectral and spatial graph convolutional networks rely on predefined graph topologies that cannot adapt to the input graph signals. In this subsection, we demonstrate that the spatial transformer can be formulated as an iterative feature learning process of message passing and update for all the nodes $v\in\mathcal{V}$ in a dynamical graph neural network. Let us denote $x_v\in {\mathbb R}^{d_{\mathcal G}}$ the input features of node $v$. For arbitrary $v\in\mathcal{V}$, it receives the message $m_v\in {\mathbb R}^{d_{\mathcal G}}$ from the nodes in $\mathcal{V}$.  
\begin{equation}\label{eq:message}
m_v=\sum_{u\in \mathcal V }F(x_v, x_u)=\sum_{u\in \mathcal V }\langle (W_q^{\mathcal S})^Tx_v,(W_k^{\mathcal S})^Tx_u\rangle x_u
\end{equation}
where $F$ is the composite message-passing function that realizes Eqs.~\eqref{eqa}, ~\eqref{eq4} and \eqref{eq5}. When $m_v$ is obtained, $x_v$ is updated with $y_v$ calculated from $m_v$ and $x_v$.
\begin{equation}\label{eq:update}
y_v=G(m_v ,x_v)
\end{equation}
In the spatial transformer, $G$ is the shared position-wise feed forward network defined in Eq.~\eqref{eq6}. Comparing Eq.~\eqref{eq:spatial} with Eqs.~\eqref{eq:message} and \eqref{eq:update}, the spatial transformer can be viewed as a general message-passing dynamical graph neural network.


\begin{figure}[!t]
\renewcommand{\baselinestretch}{1.0}
\centering
\includegraphics[width=0.5\textwidth]{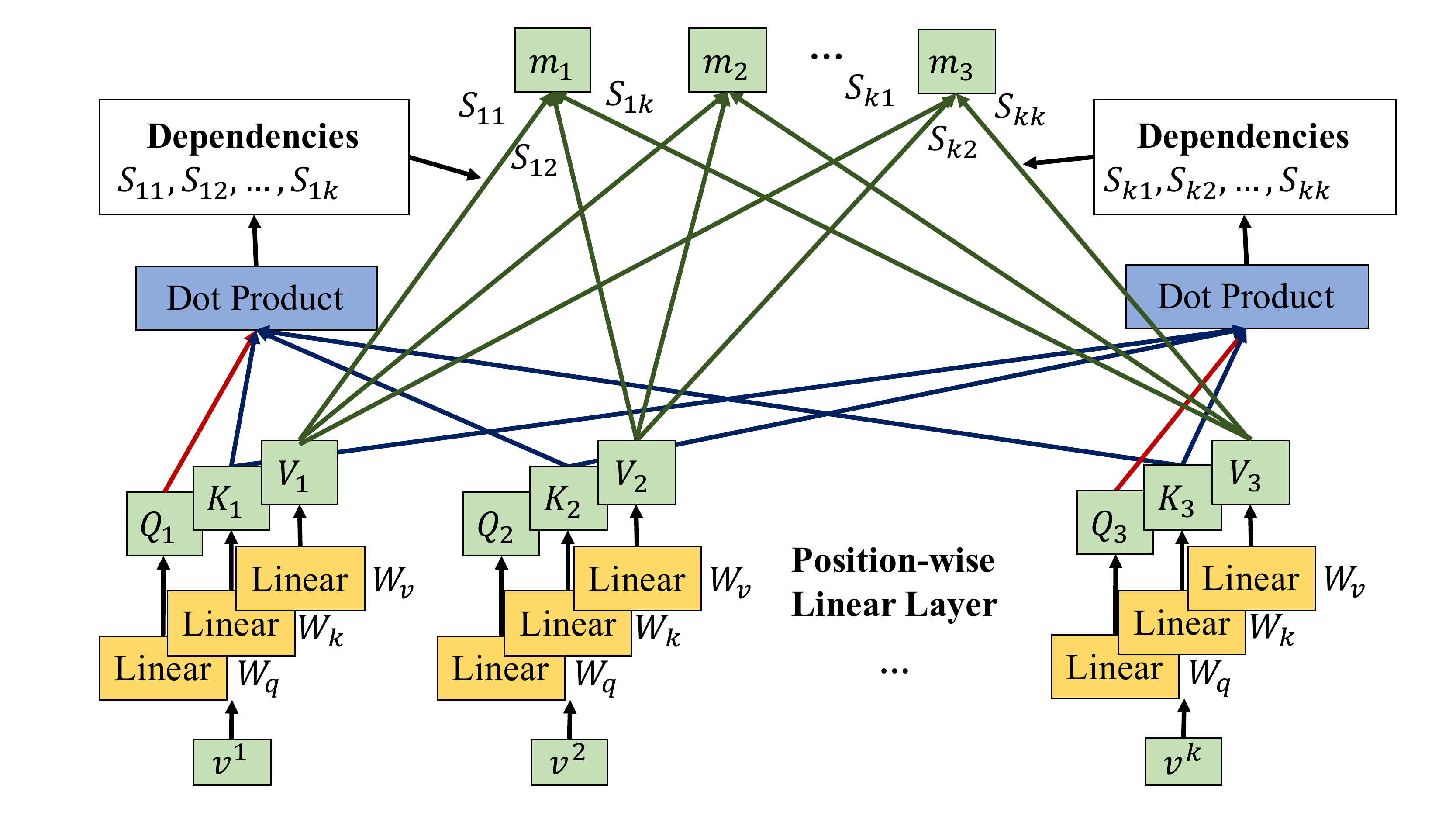}
\caption{Self-attention mechanism for long-range temporal dependencies.}\label{fig3}
\end{figure}

\subsection{Temporal Transformers}\label{sub:temporal}
Fig.~\ref{fig2}(c) depicts the proposed temporal transformer for efficiently capturing the long-range temporal dependencies~\cite{NIPS2017_7181}. In comparison to RNNs and their variants, temporal transformer can be easily scaled to long sequences with parallel processing of long-range dependencies.
Similar to the spatial transformer, ${X'^{\mathcal T}=G_t([X^{\mathcal T}, D^{\mathcal T}]}) \in {\mathbb R}^{M\times N\times d_{\mathcal G}}$ is obtained from the concatenation of the input features $X^{\mathcal T}=X^{\mathcal S}+Y^{\mathcal S}\in {\mathbb R}^{M\times N\times d_{\mathcal G}}$ and the temporal embedding ${\mathcal { D}^{\mathcal T}}$, where $G_t$ is a $1\times 1$ convolutional layer that yields a $d_{\mathcal G}$-dimensional vector for each node at each time step. Here, we also parallelize over the nodes to model temporal dependencies. The 2-D tensor of spatial features $\hat X^{\mathcal T}\in {\mathbb R}^{M\times d_{\mathcal G}}$ is considered for arbitrary one node in $\mathcal{G}$.

Self-attention mechanism is also adopted to model temporal dependencies. The input to the temporal transformer is a temporal sequence $\hat X^{\mathcal T}\in {\mathbb R}^{M\times d_{\mathcal G}}$ with a slide window of length $M$ and $d_{\mathcal G}$ channels. Similar to spatial transformer, temporal dependencies are dynamically computed in high-dimensional latent subspaces, including the query subspace spanned by $Q^{\mathcal T}\in {\mathbb R}^{M\times d^{\mathcal T}_{\mathcal A}}$, the key subspace by $K^{\mathcal T}\in {\mathbb R}^{M\times d^{\mathcal T}_{\mathcal A}}$ and the value subspace by $V^{\mathcal T}\in {\mathbb R}^{M\times d_{\mathcal G}}$.
\begin{align}
Q^ {\mathcal T}&=\hat X^{{\mathcal T}}W^{\mathcal T}_q\nonumber\\
K^{\mathcal T}&=\hat X^{\mathcal T}W^{\mathcal T}_k\nonumber\\
V^{\mathcal T}&=\hat X^{\mathcal T}W^{\mathcal T}_v,
\end{align}
where $W_q\in {\mathbb R}^{d_{\mathcal G}\times d^{\mathcal T}_{\mathcal A}}$,$W_k\in {\mathbb R}^{d_{\mathcal G}\times d^{\mathcal T}_{\mathcal A}}$ and $W_v\in {\mathbb R}^{d_{\mathcal G}\times d_{\mathcal G}}$ are the learned liner mappings.
According to Eq.~\eqref{eq1}, multi-step prediction for $v^{\tau+1},\cdots,v^{\tau+T}$ is simultaneously made from the historical observations $v^{\tau-M+1},\cdots,v^{\tau}$. We introduce the scaled dot product function to consider bi-directional temporal dependencies within $v^{\tau-M+1},\cdots,v^{\tau}$.
\begin{equation}
S^{\mathcal T}=\text{softmax}(Q^{\mathcal T}(K^{\mathcal T})^T/\sqrt{d^{\mathcal T}_{\mathcal A}})
\end{equation}
 RNN-based models are limited to consider temporal dependencies based on preceding time steps, as shown in \cite{devlin2018bert}, this left-to-right architecture is sub-optimal to model context dependencies. We further aggregate the values $V^{\mathcal T}$ with the weights $S^{\mathcal T}$ for temporal features $M^{\mathcal T}$.
\begin{equation}
M^{\mathcal T}=S^{\mathcal T}V^{\mathcal T}
\end{equation}
Fig.~\ref{fig3} illustrates the modeling of temporal dependencies.

To explore the interactions among latent features, a shared three-layer feed-forward neural network is developed for $M^{\mathcal T}$.
\begin{equation}
U^{\mathcal T}=\text{ReLu}(\text{ReLu}(M'^{\mathcal T}W^{\mathcal T}_0)W^{\mathcal T}_1)W^{\mathcal T}_2
\end{equation}
Here, the residual connection $M'^{\mathcal T}=M^{\mathcal T}+X^{\mathcal T}$ is adopted for stable training. For each node, its output is $\hat Y^{\mathcal T}=U^{\mathcal T}+M'^{\mathcal T}$.
As a result, the output of temporal transformer is $Y^{\mathcal T}\in {\mathbb R}^{M\times N\times d_{\mathcal G}}$ by collecting $\hat Y^{\mathcal T}$ for all the nodes.

Long-range bidirectional temporal dependencies are efficiently captured in each layer of the temporal transformer, as each time step attends to the remaining time steps within the slide window. The temporal transformer can be easily scaled to long sequences by increasing $M$ without much sacrifice in computation efficiency. By contrast, RNN-based models would suffer from vanishing or exploding gradients, while convolution-based models have to explicitly specify the number of convolutional layers that grows with $M$.

\section{Experiments}\label{sec:exp}
We demonstrate that STTN achieves the state-of-the-art performance in traffic flow forecasting, especially for long-term predictions, on two real-world datasets, i.e., PeMSD7(M) and PEMS-BAY. Furthermore, ablation studies have been made to validate the multi-step prediction and the effectiveness of spatial and temporal transformers for long-term traffic forecasting. We also analyze the model configurations, including the number of blocks, feature channels and layers, attention heads for the self-attention mechanism and spatial-temporal positional embedding.

\subsection{Dataset and Data Preprocessing}
Two real-world traffic datasets are adopted for evaluations:\\ 
\noindent\textbf{PeMSD7(M):} Traffic data from 228 sensor stations in the California state highway system during the weekdays from May through June in 2012. \\
\noindent\textbf{PEMS-BAY:} 6-month traffic data collected from 325 sensors in the Bay Area of California, starting from January 1st 2017 through May 31th 2017.

Traffic speeds are aggregated every five minutes and normalized with $Z$-Score as inputs. The road topology information is represented by a graph adjacency matrix. The graph of PeMS-BAY dataset is pre-designed as a directed graph to differentiate the influence of different directions. In~\cite{li2018dcrnn_traffic}, forward and backward diffusion graph convolutions are adopted to model the directed spatial dependencies. However, it is difficult to construct a directed graph with a proper  metric for the influence of directions.
In this paper, we use the self-attention mechanism to model directed spatial dependencies in a data-driven manner and alleviate the computation burden on differentiating the influence of directions. Only the undirected graph adjacency matrix is required to represent the distance and connectivity between sensors. In PEMS-BAY, the undirected graph for road topology is generated by selecting the larger weight from the two directions (i.e., upstream and downstream) of each pair of nodes. The adjacency matrix is symmetric based on the distances between sensors in PeMSD7.

\subsection{Experimental Settings}
All experiments are conducted on a NVIDIA 1080Ti GPU. The proposed model is trained with the mean absolute error (MAE) loss using the RMSprop optimizer for 50 epochs with a batch size of 50. The initial leaning rate is set to $10^{-3}$ and decays at a rate of 0.7 for every five epochs. Table~\ref{table1} shows the average results for STTN obtained by five independent trials of the same experiment on each dataset. For evaluations, we adopt the results reported in~\cite{yu2018spatio} and~\cite{li2018dcrnn_traffic}, where 12 current observations (60 minutes) are used to predict the traffic conditions in the next 15, 30 and 45 minutes in PeMSD7(M) and 15, 30 and 60 minutes in PEMS-BAY, respectively. Graph WaveNet~\cite{wu2019graph} is trained on PeMSD7(M) using its the publicly released code and the best performance is reported in Table~\ref{table1}.

\begin{table*}[!t]
\renewcommand{\baselinestretch}{1.0}
\renewcommand{\arraystretch}{1.0}
\centering
\caption{MAE, MAPE (\%) and RMSE for PEMS-BAY and PeMSD7(M) obtained by STTN and the baselines. Traffic conditions in the next 15, 30 and 45 minutes are predicted for PeMSD7(M) and 15, 30 and 60 minutes for PEMS-BAY.}\label{table1}
\begin{tabular}{c||c|c|c||c|c|c}
\hline\hline
\multirow{2}*{Model}&\multicolumn{3}{c||}{PEMS-BAY (15/30/60 min)}&\multicolumn{3}{c}{PeMSD7(M) (15/30/45 min)}\\
\cline{2-7}
&MAE&MAPE (\%)&RMSE&MAE&MAPE (\%)&RMSE\\
\hline
HA&2.88&6.8&5.59&4.01&10.61&7.20\\
\hline
ARIMA~\cite{makridakis1997arma}&1.62/2.33/3.38&3.5/5.4/8.3&3.30/4.76/6.50&5.55/5.86/6.27&12.92/13.94/15.20&9.00/9.13/9.38\\
\hline
LSVR~\cite{wu2004travel}&1.85/2.48/3.28&3.8/5.5/8.0&3.59/5.18/7.08&2.50/3.63/4.54&5.81/8.88/11.50&4.55/6.67/8.28\\
\hline
FNN&2.20/2.30/2.46&5.19/5.43/5.89&4.42/4.63/4.89&2.74/4.02/5.04&6.38/9.72/12.38&4.75/6.98/8.58\\
\hline
FC-LSTM~\cite{Sutskever2014-NIPS}&2.05/2.20/2.37&4.8/5.2/5.7&4.19/4.55/4.96&3.57/3.94/4.16&8.60/9.55/10.10&6.20/7.03/7.51\\
\hline
DCRNN~\cite{li2018dcrnn_traffic}&1.38/1.74/2.07&2.9/3.9/4.9&2.95/3.97/4.74&2.37/3.31/4.01&5.54/8.06/9.99&4.21/5.96/7.13\\
\hline
STGCN~\cite{yu2018spatio}&1.39/1.84/2.42&3.00/4.22/5.58&2.92/4.12/5.33&2.25/3.03/3.57&5.26/7.33/8.69&4.04/5.70/6.77\\
\hline
Graph WaveNet~\cite{wu2019graph} &1.30/1.63/1.95&2.74/3.70/4.52&2.73/3.67/4.63&{2.14}/2.80/3.19&{4.93}/6.89/8.04&{4.01}/5.48/6.25\\
\hline
STTN&\bf 1.36/1.67/1.95& \bf2.89/3.78/4.58&\bf2.87/3.79/4.50&\bf2.14/2.70/3.03&\bf5.05/6.68/7.61&\bf4.04/5.37/6.05\\
\hline\hline
\end{tabular}
\end{table*}

\begin{figure*}[!t]
\renewcommand{\baselinestretch}{1.0}
\centering
\subfloat[STTN]{\includegraphics[width=0.5\textwidth]{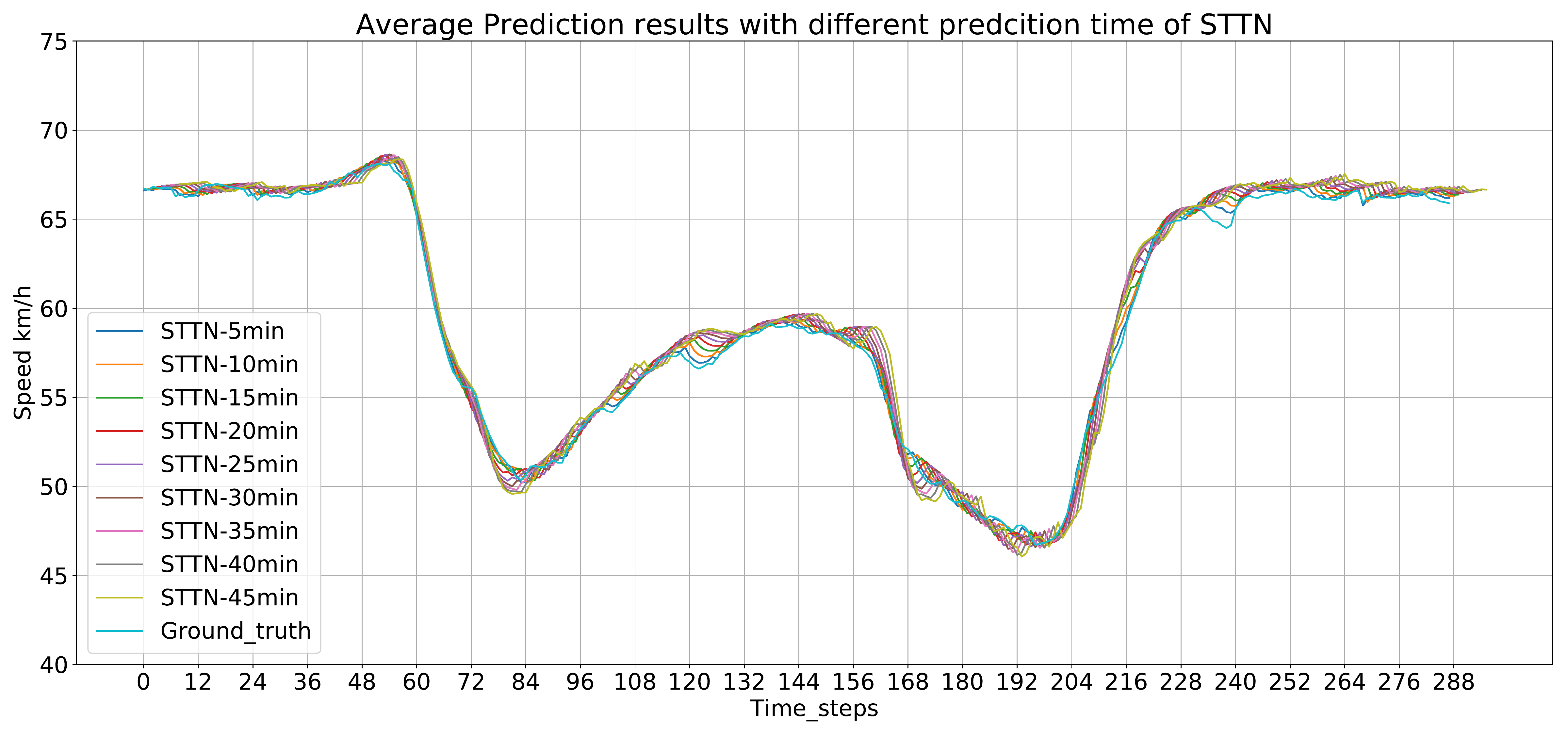}}
\subfloat[Graph WaveNet~\cite{wu2019graph}]{\includegraphics[width=0.5\textwidth]{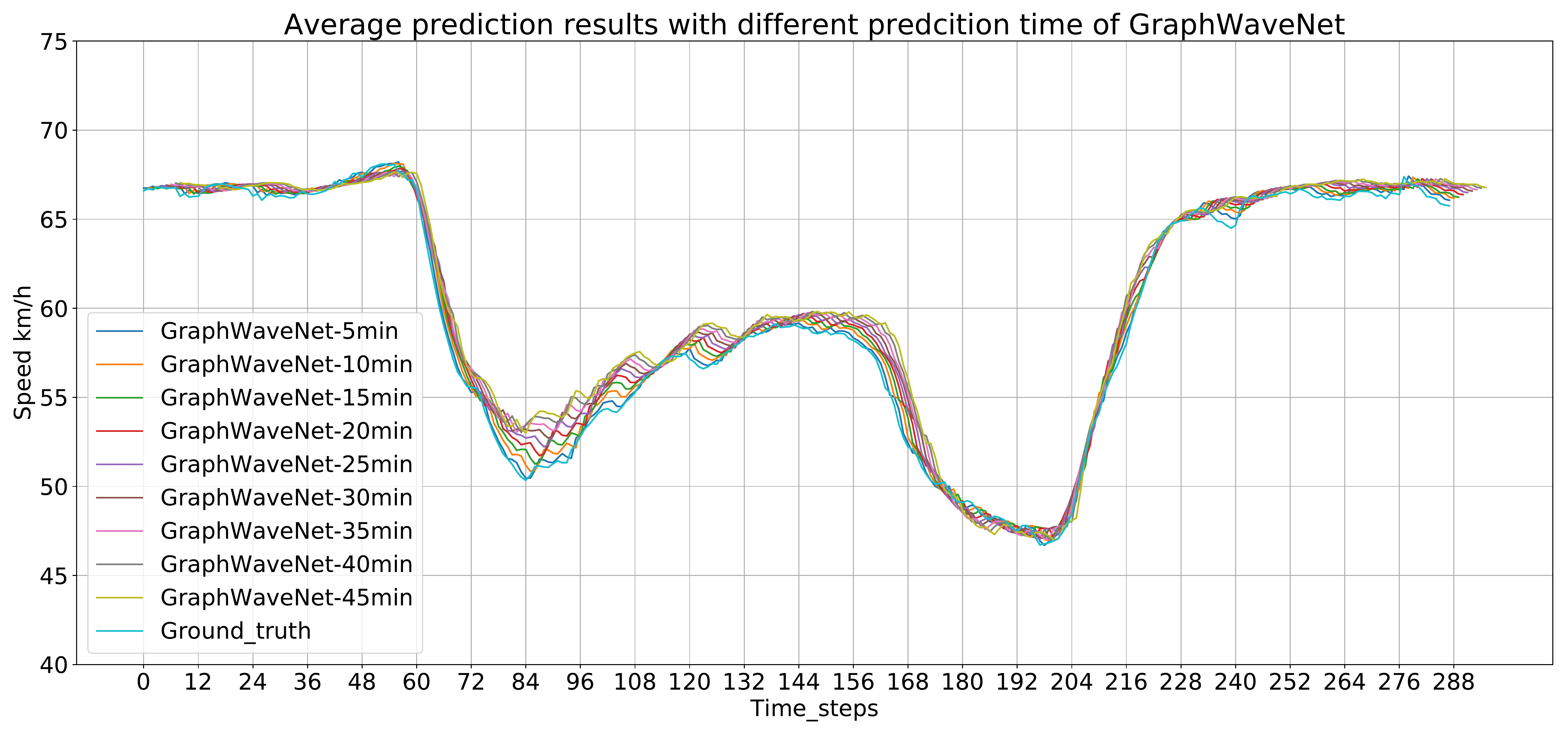}}\\
\subfloat[STGCN~\cite{yu2018spatio}]{\includegraphics[width=0.5\textwidth]{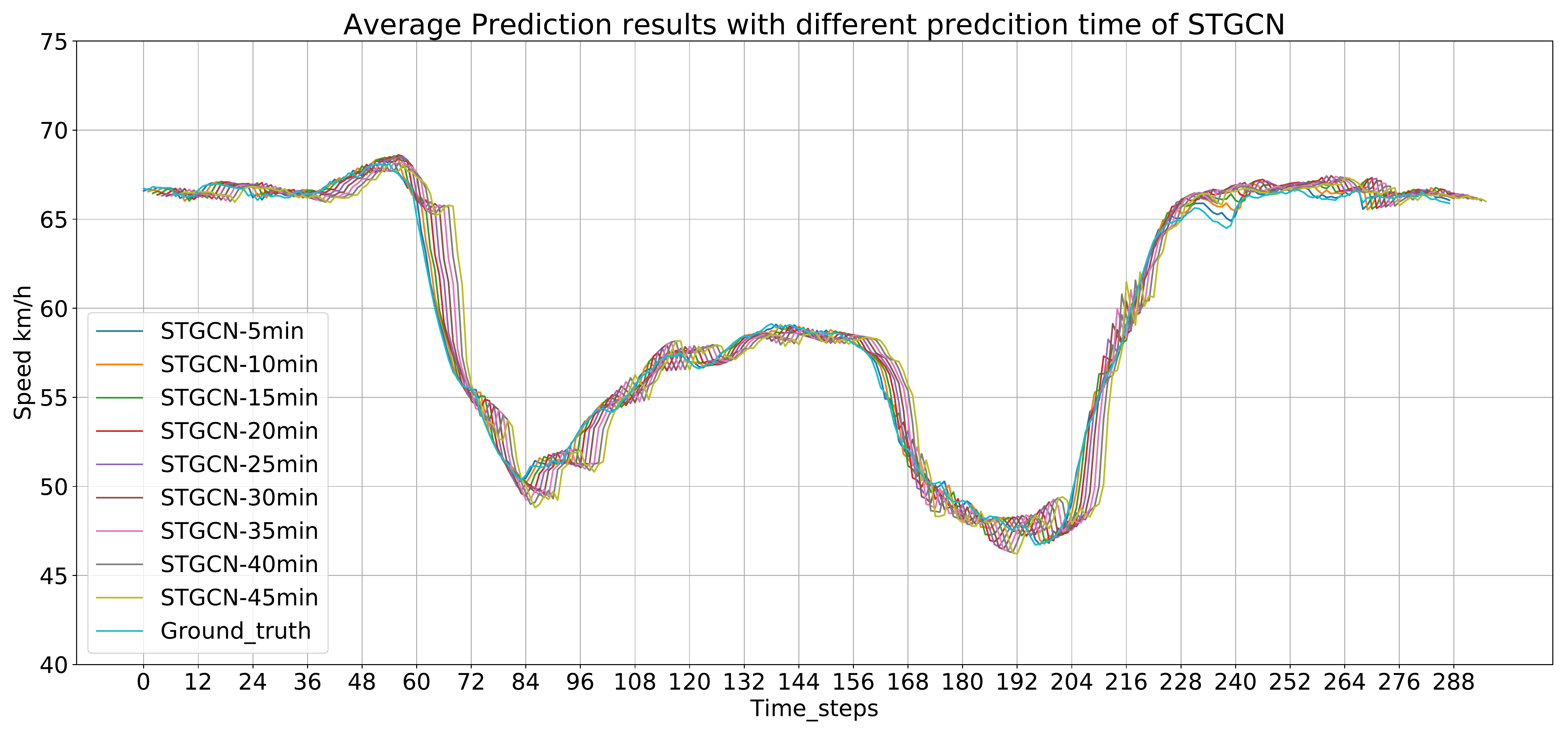}}
\subfloat[DCRNN~\cite{li2018dcrnn_traffic}]{\includegraphics[width=0.5\textwidth]{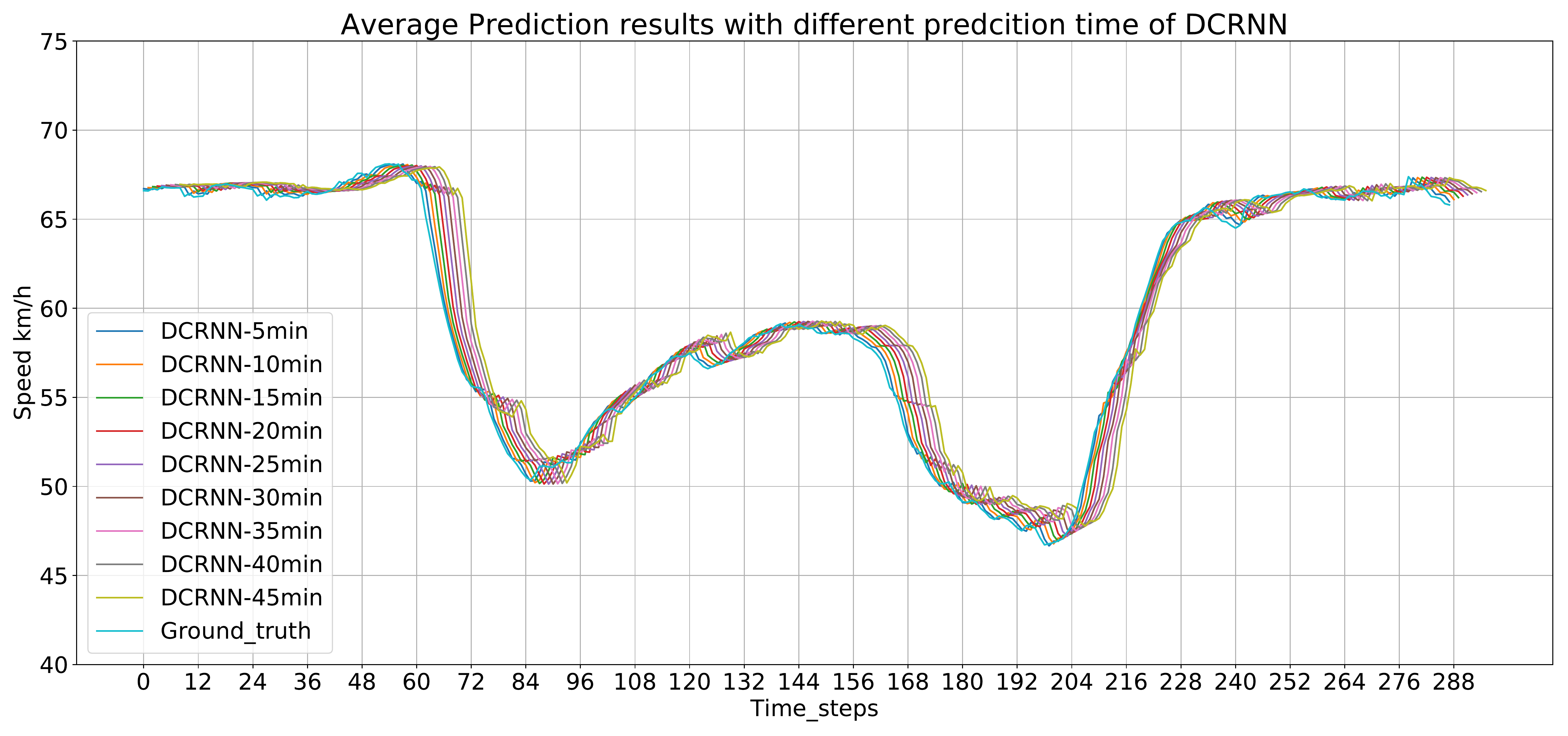}}\\
\caption{Visualization of one-day traffic flow forecasting for PeMSD7 obtained by STTN, Graph WaveNet~\cite{wu2019graph}, STGCN~\cite{yu2018spatio} and DCRNN~\cite{li2018dcrnn_traffic}.}
\label{fig4}
\end{figure*}

\subsection{Evaluation Metrics and Baselines}
We evaluate STTN and benchmark traffic forecasting methods in terms of mean absolute error (MAE), mean absolute percentage error (MAPE) and root mean squared error (RMSE). Baselines include historical average (HA), autoregressive integrated moving average (ARIMA) with Kalman filtering~\cite{makridakis1997arma}, linear support vector regressiion (LSVR)~\cite{wu2004travel}, feed-forward neural network (FNN), fully-connected LSTM (FC-LSTM)~\cite{Sutskever2014-NIPS}, STGCN~\cite{yu2018spatio}, DRCNN~\cite{li2018dcrnn_traffic} and Graph WaveNet \cite{wu2019graph} .

For PeMSD7(M), one spatial-temporal block is adopted. Two hidden layers and one single attention are adopted for each spatial and temproal transformer with 64 feature channels. Considering that PeMS-BAY is much larger than PeMSD7(M) in spatial and temporal scale, three spatial-temporal blocks are stacked to model spatial-temporal dependencies. In each spatial-temporal block, each spatial and temporal transformer consists of one hidden layer and single attention with 64 feature channels. Residual structures are adopted for stable learning and fast convergence.

\subsection{Experimental Results}
Table~\ref{table1} provides MAE, MAPE and RMSE for STTN and baselines for traffic forecasting with varying period of time steps on PEMS-BAY and PeMSD7(M). 

\noindent\textbf{PeMSD7(M):} STTN outperforms STGCN~\cite{yu2018spatio} and DCRNN~\cite{li2018dcrnn_traffic} by a large margin that grows with the range of time steps for prediction. In comparison to Graph WaveNet~\cite{wu2019graph}, STTN performs better in long-term prediction ($\ge$30 minutes) and yields competitive performance for short-term prediction ($\le$30 minutes). These facts imply that long-term prediction can be facilitated by jointly considering dynamic spatial dependencies and long-range temporal dependencies. By contrast, Graph WaveNet leverages convolutional kernels with small receptive fields to capture stationary spatial-temporal dependencies for short-term prediction. 

\noindent\textbf{PEMS-BAY:} STTN is competitive with Graph WaveNet and outperforms STGCN and DCRNN. Compared with STGCN, Graph WaveNet and DCRNN employ bi-directional diffusion graph convolutions based on the non-symmetric adjacency matrix explicitly designed for the influence of directions. STTN leverages the self-attention mechanism to learn dynamical directed spatial dependencies from the symmetric adjacency matrix (without prior information for upstream/downstream traffic flows). It is worth mentioning that, compared with STGCN that also consists of three spatial-temporal blocks, STTN improves the prediction performance considerately.

For further evaluation, traffic forecasting for one-day period is visualized in for STNN, Graph WaveNet, STGCN and DCRNN. Here, the one-day prediction is obtained for each time step by averaging along the spatial dimension on test dataset of PeMSD7(M). Fig.~\ref{fig4} shows that STTN and Graph WaveNet improve traffic flow forecasting in changing area, e.g. $\tau\in[60,84]$, in comparison to STGCN and DCRNN. Note that time shifts of the curves of predictions are evident for STGCN and DCRNN, which suggests that prediction errors grow with the time steps, especially in the areas with sharp variation. Moreover, STTN can capture continuous changes in a long period of time steps, e.g. $\tau\in[84,192]$. This fact implies that dynamical spatial dependencies and long-range temporal dependencies captured by STTN benefit traffic flow forecasting, especially long-term prediction.

\subsection{Computational Complexity}
We further evaluate the computational costs for DCRNN, STGCN, Graph WaveNet and STTN. All the experiments are conducted on the same GPU. Table~\ref{table6} reports the average training speed for one epoch. STGCN is efficient with the fully convolutional structures. DRCNN is time-consuming due to the recurrent structures for training with joint loss for multiple time steps, as its training time is proportional to the number of prediction time steps. STTN yields a reduction of 10-40\% and 40-60\% in computational costs in comparison to Graph WaveNet and DCRNN, respectively. Note that STTN is scalable to achieve long-term prediction without excessive computational complexity.
\begin{table}[!t]
\renewcommand{\baselinestretch}{1.0}
\renewcommand{\arraystretch}{1.0}
\caption{Average training time (sec/epoch) for PEMS-BAY and PeMSD7(M) obtained by STTN, Graph WaveNet, STGCN and DRCNN, respectively.}\label{table6}
\centering
\begin{tabular}{c||c|c|c|c}
\hline \hline
\multirow{2}*{Dataset}&\multicolumn{4}{c}{Average training time (sec/epoch)}\\
\cline{2-5}
&STTN&Graph WaveNet&STGCN&DRCNN\\
\hline\hline
PEMS-bay&458&507&99&809\\
\hline
PeMSD7(M)&45&72&10&108\\
\hline\hline
\end{tabular}
\end{table}

\subsection{Ablation Studies}
Ablation studies have been made on the PeMSD7(M) dataset to verify the design of STTN. Here, PeMSD7(M) is selected, as it is much more challenging than PEMS-BAY in the sense of smaller scale and complex sptial-temporal dependencies. For example, traffic speeds in PeMSD7(M) have a larger standard deviation than those in PEMS-BAY. For effective evaluation, we use only one spatial-temporal block with 64 feature channels for the spatial and temporal transformer.

\subsubsection{Multi-step Prediction vs. Autoregressive Prediction}
\begin{table}[!t]
\renewcommand{\baselinestretch}{1.0}
\renewcommand{\arraystretch}{1.0}
\caption{MAE, MAPE (\%) and RMSE for PeMSD7(M) obtained by STGCN and STTN with autoregressive (AR) and multi-step (MS) prediction.}\label{table2}
\centering
\begin{tabular}{c|c||c|c|c}
\hline \hline
\multicolumn{2}{c||}{\multirow{2}*{Model}}&\multicolumn{3}{c}{PeMSD7(M) (15/30/45 min)}\\
\cline{3-5}
\multicolumn{2}{c||}{}&MAE&MAPE (\%)&RMSE\\
\hline
\multirow{2}*{STGCN}&AR&2.25/3.03/3.57&5.26/7.33/8.69&4.04/5.70/6.77\\
\cline{2-5}
&MS&2.25/2.90/3.32&5.33/7.19/8.42&4.19/5.59/6.42\\
\hline
\multirow{2}*{STTN}&AR&2.19/2.97/3.54&5.10/7.20/8.75&4.17/5.89/7.05\\
\cline{2-5}
&MS&2.14/2.75/3.12&5.06/6.84/7.89&4.03/5.41/6.17\\
\hline \hline
\end{tabular}
\end{table}

Autoregressive prediction is prevailing in traffic flow forecasting, in which prediction for each time step is leveraged for the succeeding predictions. However, autoregressive prediction would cause error prediction, due to the accumulated error for step-by-step prediction. Thus, it would hamper long-term prediction. DCRNN~\cite{li2018dcrnn_traffic} develops a sampling scheme to address this problem. In this paper, we explicitly make long-term multi-step prediction from the historical observations, rather than based on the predicted values. For validation, we compare STGCN~\cite{yu2018spatio} and STTN with one ST block in both autoregressive and multi-step prediction. Table~\ref{table2} shows that STTN yields noticeable gains in MAE, MAPE and RMSE in comparison to STGCN and STTN with autoregressive prediction. It should be noted that prediction error grows slowly for STTN in long-term prediction, when compared with the other models.

\begin{table}[!t]
\renewcommand{\baselinestretch}{1.0}
\renewcommand{\arraystretch}{1.0}
\caption{MAE, MAPE (\%) and RMSE for PeMSD7(M) obtained by STTN with fixed graph convolution (Baseline), STTN with attention to local nodes (STTN-S (local)) and STTN using the spatial transformer with $a$ attention heads and $h$ hidden layers (STTN-S$(a,h)$).}\label{table3}
\centering
\begin{tabular}{c||c|c|c}
\hline \hline
\multirow{2}*{Model}&\multicolumn{3}{c}{PeMSD7(M) (15/30/45 min)}\\
\cline{2-4}
&MAE&MAPE (\%)&RMSE\\
\hline
Baseline&2.18/2.92/3.43&5.12/7.18/8.65&4.04/5.57/6.58\\
\hline
STTN-S (local)&2.18/2.86/3.30&5.15/7.05/8.33&4.04/5.48/6.36\\
\hline
STTN-S(1,1)&2.16/2.79/3.16&5.09/6.92/8.00&4.04/5.42/6.19\\
\hline
STTN-S(2,1)&2.15/2.77/3.12&5.09/6.88/7.91&4.01/5.37/6.12\\
\hline
STTN-S(4,1)&2.14/2.75/3.09&5.03/6.78/7.79&4.00/5.35/6.08\\
\hline
STTN-S(1,2)&2.14/2.75/3.08&5.00/6.69/7.63&4.01/5.36/6.08\\
\hline
STTN-S(1,3)&2.15/2.75/3.08&5.07/6.80/7.75&4.02/5.39/6.09\\
\hline \hline
\end{tabular}
\end{table}
\begin{figure*}[!t]
\renewcommand{\baselinestretch}{1.0}
\centering
\subfloat[Short-term prediction (5 minutes)]{\includegraphics[width=0.5\textwidth]{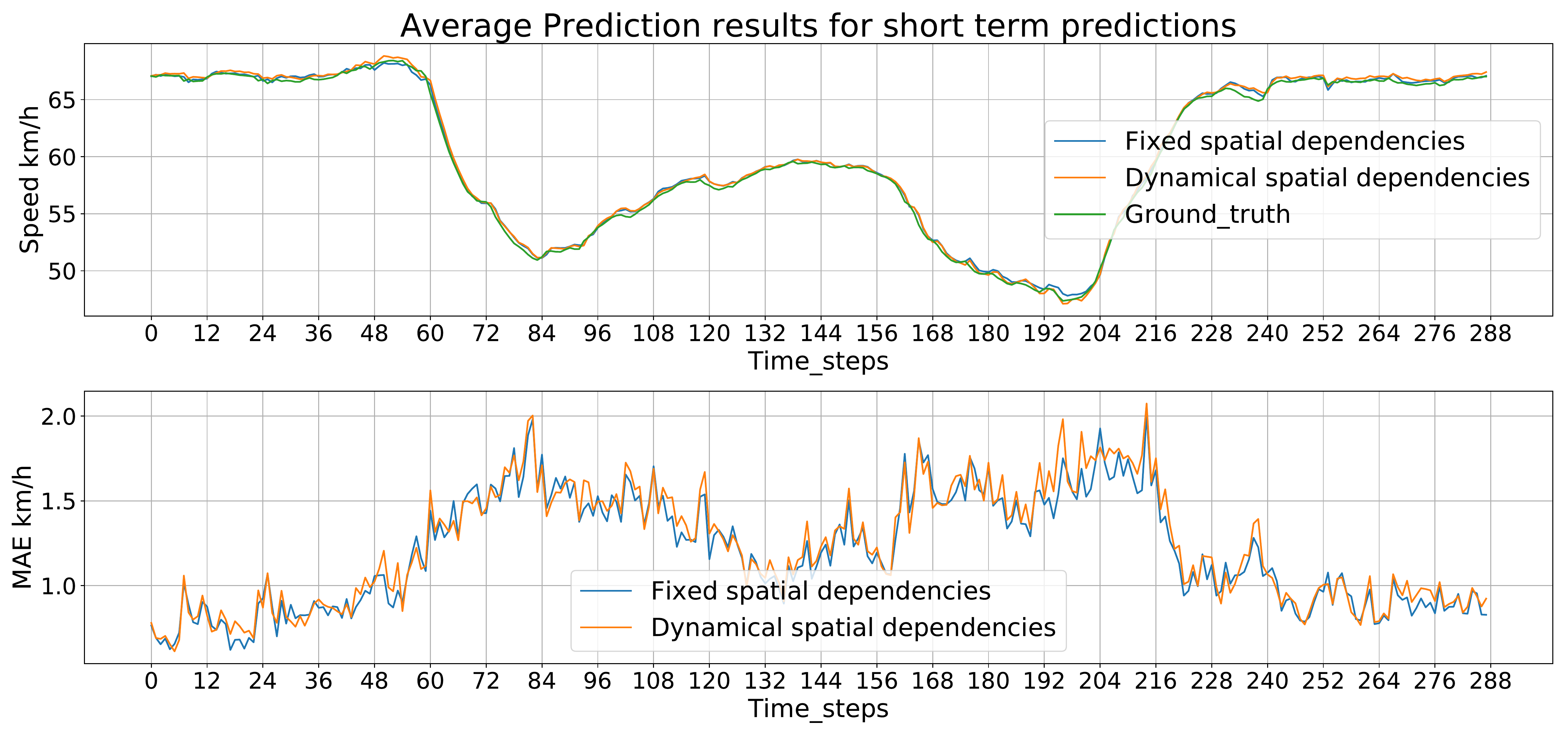}}
\subfloat[Long-term prediction (60 minutes)]{\includegraphics[width=0.5\textwidth]{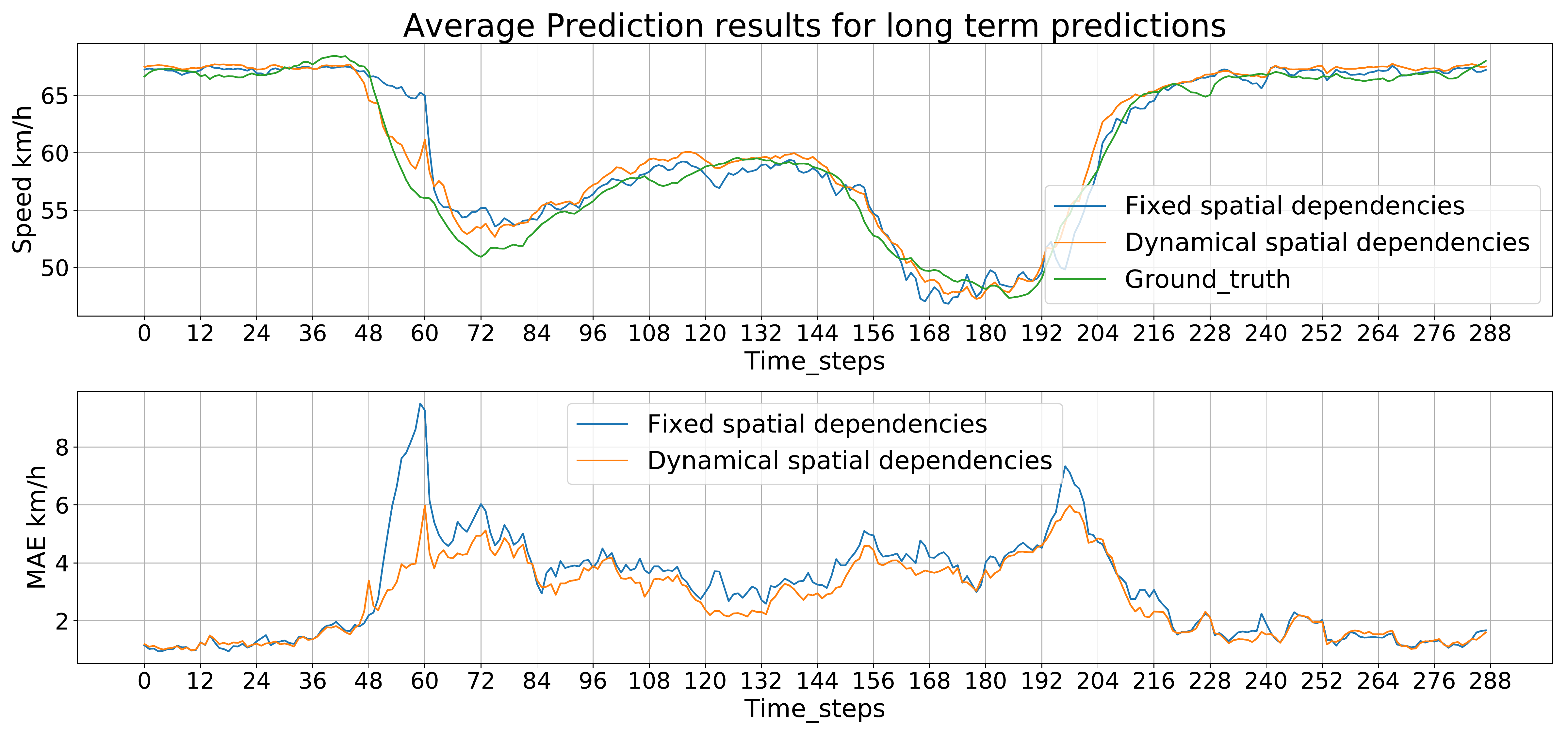}}\\
\caption{Average traffic speeds and MAE for one-day traffic flow forecasting using short-term and long-term traffic prediction with the baseline (fixed spatial dependencies) and STTN-S(1,1) (dynamical spatial dependencies).}\label{fig5}
\end{figure*}

\subsubsection{Effectiveness of Spatial Transformer}\label{sub:abl-spatial}

We demonstrate that the proposed spatial transformer can model dynamical spatial dependencies to improve the performance of long-term prediction. Variants of STTN are considered to evaluate the methods for modeling spatial dependencies. The baseline consists of one fixed graph convolutional layer realized by Chebyshev polynomial approximation and one convolution-based sequence modeling module (GLU) adopted in STGCN \cite{yu2018spatio}. Similar to ~\cite{pan2019urban}, STTN-S (local) is the attention-based method limited to the $k$-nearest neighboring nodes by masking the learned matrix for dynamical dependencies. STTN-S$(a,h)$ stands for STTN using the proposed spatial transformer with $a$ attention heads and $h$ hidden layers. Consequently, the baseline only models fixed spatial dependencies, while STTN-S (local) and STTN-S$(a,h)$ consider local and global dynamical spatial dependencies, respectively.

\begin{figure*}[t]
\centering
\includegraphics[width=0.99\textwidth]{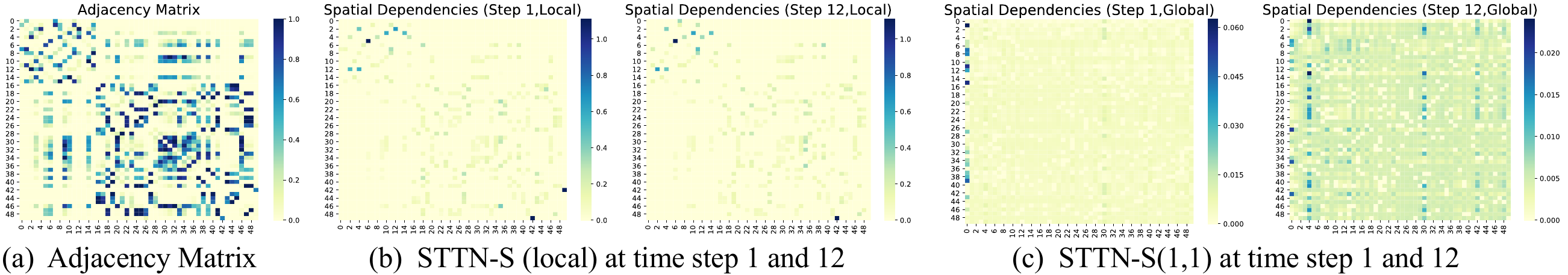}
\caption{Spatial dependencies learned for the first 50 sensors in PeMSD7(M). The adjacency matrix only keeps local nodes. 
}\label{fig7}
\end{figure*}

Table~\ref{table3} shows that STTN-S$(1,1)$ outperforms the baseline by a large margin, especially for long-term prediction. This fact implies that the spatial transformer can exploit dynamical spatial dependencies to achieve accurate long-term prediction. Fig.~\ref{fig5} illustrates the averaged results for one-day traffic flow forecasting with short-term (5-min) and long-term (60-min) prediction on the test dataset for the baseline and STTN-S$(1,1)$. STTN-S$(1,1)$ achieves better performance for long-term predictions, especially in sharply changing areas, e.g., the period of time steps [48,84] in Fig.~\ref{fig5}(b). 
We further evaluate the spatial transformers that capture local and global spatial dependencies. 
According to Table~\ref{table3}, STTN-S (local) with the local constraint is inferior to STTN with the proposed spatial transformer. This fact suggests that global dynamical spatial dependencies can facilitate the traffic flow forecasting, when compared with local dependencies. We also compare the learned spatial dependencies for STTN-S (local) and STTN-S$(1,1)$ in Fig.~\ref{fig7}. Fig.~\ref{fig7}(c) shows that the spatial dependencies STTN-S (local) relates the sensors within a local neighborhood, while STTN-S$(1,1)$ increase with the growth of time steps, due to the small distances between most adjacent sensors. 

Furthermore, Table~\ref{table3} provides the MAE, MAPE and RMSE obtained under various numbers of attention heads and hidden layers in spatial transformer. Traffic flow forecasting performance is continuously improved when increasing attention heads, as multi-head attention can model spatial dependencies in different latent subspaces to further utilize hidden patterns of dependenceis. On the contrary, a larger number of hidden layers would trivially benefit the performance. This fact implies that one hidden layer would be enough to capture the spatial dependencies for relatively small PeMSD7(M).

\begin{table}[!t]
\renewcommand{\baselinestretch}{1.0}
\renewcommand{\arraystretch}{1.0}
\caption{MAE, MAPE (\%) and RMSE for PeMSD7(M) obtained by the fixed graph convolution with convolution kernel sizes 3 (Baseline), 6 (Conv-6), 9 (Conv-9) and 12 (Conv-12) and STTN using the temporal transformer with $a$ attention heads and $h$ hidden layers (STTN-T$(a,h)$).}\label{table4}
\centering
\resizebox{0.5\textwidth}{!}{
\begin{tabular}{c||c|c|c}
\hline \hline
Model&MAE&MAPE (\%)&RMSE\\
\hline
Baseline&2.18/2.92/3.43&5.12/7.18/8.65&4.04/5.57/6.58\\
\hline
Conv-6&2.16/2.87/3.35&5.07/7.05/8.42&4.01/5.49/6.44\\
\hline
Conv-9&2.16/2.87/3.34&5.08/7.08/8.48&4.02/5.50/6.43\\
\hline
Conv-12&2.16/2.85/3.31&5.07/7.02/8.36&4.00/5.45/6.37\\
\hline
\hline
STTN-T(1,1)&2.19/2.88/3.36&5.15/7.10/8.46&4.08/5.53/6.46\\
\hline
STTN-T(2,1)&2.19/2.88/3.36&5.15/7.10/8.46&4.08/5.53/6.46\\
\hline
STTN-T(4,1)&2.18/2.89/3.37&5.14/7.10/8.48&4.07/5.53/6.47\\
\hline
STTN-T(1,2)&2.17/2.86/3.32&5.10/7.05/8.40&4.05/5.50/6.43\\
\hline
STTN-T(1,3)&2.16/2.86/3.32&5.09/7.07/8.42&4.03/5.50/6.43\\
\hline\hline
\end{tabular}
}
\end{table}
\subsubsection{Effectiveness of Temporal Transformer}
\begin{figure}[!t]
\renewcommand{\baselinestretch}{1.0}
\centering
\includegraphics[width=0.48\textwidth]{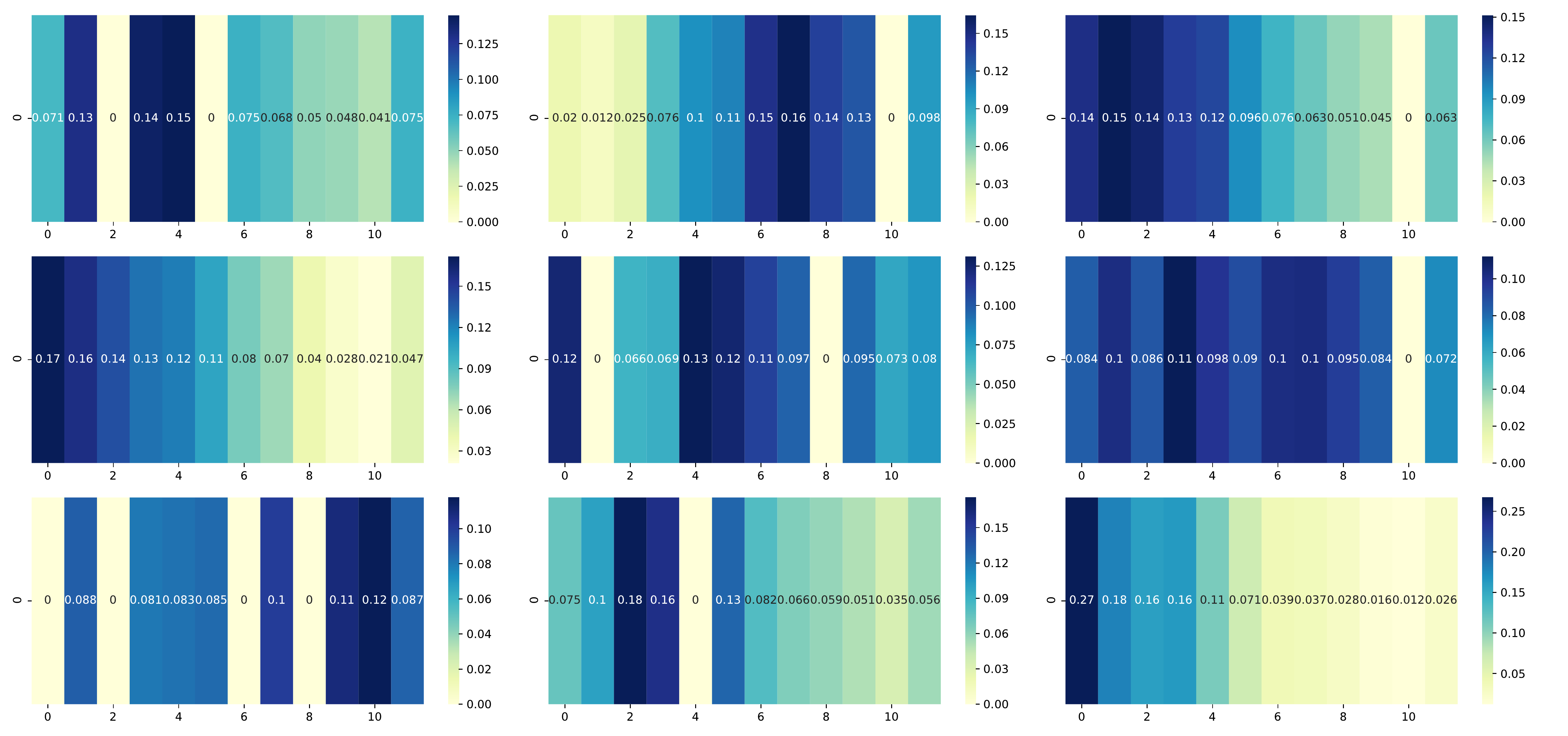}\\
\caption{Attention matrices of the first nine sensors for the temporal transformer generated by 15-min prediction on test dataset. Darker color indicates larger attention weights.}\label{fig6}
\end{figure}

We further validate that the proposed temporal transformer is efficient to capture long-range temporal dependencies for accurate traffic flow forecasting. The same baseline with fixed graph convolution is adopted as in Section~\ref{sub:abl-spatial}
The receptive fields of convolution kernel in GLU layer are adjusted to control the range of temporal dependencies. Here, we consider the convolution kernel size 3 (baseline), 6 (Conv-6), 9 (Conv-9) and 12 (Conv-12). 
Table~\ref{table4} shows that long-term prediction can be improved with long-range temporal dependencies determined by the large convolution kernel sizes. 
Thus, we substitute the GLU layer with the proposed temporal transformer to validate its effectiveness. Table~\ref{table4} shows that temporal transformer is superior to the fixed graph convolution in long-term prediction. In Fig.~\ref{fig6}, we further illustrate the attention matrices of the first nine sensors for the temporal transformer. The weights for temporal attention are different for different sensors. In some cases, earliest time-steps are utilized with long-range dependencies for multi-step prediction. 

The effects of the number of attention heads and hidden layers are also evaluated for the proposed temporal transformer. Table~\ref{table4} indicates that multi-head attention would not benefit the traffic flow forecasting, as temporal dependencies are not as complex as spatial dependencies. We also find that traffic flow forecasting tends to be improved by increasing the number of hidden layers. 

\subsection{Model Configurations}
\begin{table*}[!t]
\renewcommand{\baselinestretch}{1.0}
\renewcommand{\arraystretch}{1.0}
\caption{MAE for PeMSD7(M) obtained by STTN with various model configurations.}\label{table7}
\centering
\begin{tabular}{c|ccccc|c}
\hline \hline
Model&\# of &\# of &\# of hidden layers& \# of attention heads&Positional&\multirow{2}*{MAE (15/30/45 min)}\\
configurations&blocks&feature channels&$(h_{\mathcal S},h_{\mathcal T})$&$(a_{\mathcal S},a_{\mathcal T})$&embedding&\\
\hline
\multirow{3}*{Blocks}&1&[64,64]&(1,1)&(1,1)&\checkmark&2.17/2.78/3.14\\
&2&[64,64]$\times$2&(1,1)&(1,1)&\checkmark&2.13/2.71/3.04\\
&3&[64,64]$\times$3&(1,1)&(1,1)&\checkmark&2.13/2.71/3.05\\
\hline
\multirow{2}*{Channels}&1&[32,32]&(1,1)&(1,1)&\checkmark&2.18/2.82/3.21\\
&1&[128,128]&(1,1)&(1,1)&\checkmark&2.16/2.76/2.13\\
\hline
\multirow{3}*{Layers}&1&[64,64]&(1,2)&(1,1)&\checkmark&2.16/2.77/3.13\\
&1&[64,64]&(2,1)&(1,1)&\checkmark&2.15/2.75/3.08\\
&1&[64,64]&(2,2)&(1,1)&\checkmark&2.13/2.72/3.05\\
\hline
\multirow{3}*{Attention}&1&[64,64]&(1,1)&(4,1)&\checkmark&2.15/2.74/3.09\\
&1&[64,64]&(1,1)&(1,4)&\checkmark&2.15/2.75/3.11\\
&1&[64,64]&(1,1)&(2,2)&\checkmark&2.14/2.74/3.10\\
\hline
\multirow{3}*{Embeddings}&1&[64,64]&(1,1)&(1,1)&w/o S&2.18/2.84/3.26\\
&1&[64,64]&(1,1)&(1,1)&w/o T&2.17/2.79/3.16\\
&1&[64,64]&(1,1)&(1,1)&w/o ST&2.19/2.86/3.31\\
\hline\hline
\end{tabular}
\end{table*}
Finally, we discuss the model configurations for STTN, including the number of spatial-temporal blocks, the number of feature channels, the number of hidden layers, the number of attention heads and positional embedding. Table~\ref{table7} summarizes the MAE for PeMSD7(M) obtained by STTNs with various model configurations. MAE decreases by cascading multiple spatial-temporal blocks to jointly model spatial-temporal dependencies, but would be stable when enough spatial-temporal blocks are stacked (e.g., greater than 2 blocks in Table~\ref{table7}).   

In each spatial-temporal block, we investigate the effect of number of feature channels, hidden layers and attention heads. Here, the number of feature channel indicates the dimension of subspace in which dependencies are dynamically computed. The latent subspaces with higher dimensions are demonstrated to exploit more information to achieve accurate prediction. 
In comparison to the temporal transformer, traffic flow forecasting would be improved by increasing the number of hidden layers for the spatial transformer. This fact implies that long-term prediction tends to be affected by the model capacity of spatial transformer. Table~\ref{table7} also suggests that it would be better to jointly enhance the capacity of spatial and temporal transformer. 
Furthermore, multi-head attention is demonstrated to facilitate STTN in the traffic flow forecasting, especially for long-term prediction. For real-world traffic networks, relations among nodes are supposed to reside in different latent subspaces to evaluate the similarities of hidden patterns of traffic flows. This fact suggests that multi-head attention tends to be helpful for exploiting these hidden patterns, whereas the performance gain led by additional hidden patterns of spatial dependencies would be limited. 
Finally, we find that both spatial and temporal positional embedding boost the performance of traffic flow forecasting with STTN.

\section{Conclusion}\label{sec:con}
In this paper, we propose a novel paradigm of spatial-temporal transformer networks to improve the long-term prediction of traffic flows. It can dynamically model various scales of spatial dependencies as well as capture long-range temporal dependencies. Experimental results on two real-world datasets demonstrate the superior performance of the proposed STTN, especially in long-term prediction. Furthermore, the proposed spatial transformer can be generalized for dynamical graph feature learning in a variety of applications. We will further investigate this topic in future.


%





\ifCLASSOPTIONcaptionsoff
  \newpage
\fi



\bibliographystyle{IEEEtran}
\bibliography{STTN}
\end{document}